\documentclass[10pt,amsmath,amssymb,aps,prx, twocolumn,letterpaper,floatfix,superscriptaddress]{revtex4-1}
\usepackage{graphicx, xcolor} 
\usepackage{bm}
\usepackage{textcomp}
\usepackage{gensymb}
\usepackage{array,multirow}
\newcolumntype{K}[1]{>{\centering\arraybackslash}m{#1}}
\usepackage{makecell}
\usepackage{tablefootnote}
\usepackage{blindtext}

\graphicspath{{./figures/}}
\renewcommand{\(}{\left(}
\renewcommand{\)}{\right)}


\begin{document}
	\title{Unraveling Single-Particle Trajectories Confined in Tubular Networks}

	\author{Yunhao Sun$^*$}
	\affiliation{Department of Physics, University of California, San Diego, San Diego, California 92093}
	\author{Zexi Yu$^*$}
	\affiliation{Department of Physics, University of California, San Diego, San Diego, California 92093}
	
	\author{Christopher J. Obara}
	\affiliation{Janelia Research Campus, Howard Hughes Medical Institute, Ashburn, Virginia 20147}
	
	\author{Keshav Mittal}
	\affiliation{Department of Physics, University of California, San Diego, San Diego, California 92093}
	
	\author{Jennifer Lippincott-Schwarz}
	\affiliation{Janelia Research Campus, Howard Hughes Medical Institute, Ashburn, Virginia 20147}
	
	\author{Elena F. Koslover}
	\email{ekoslover@ucsd.edu}
	\affiliation{Department of Physics, University of California, San Diego, San Diego, California 92093}
	
		\date{\today}
		\preprint{}

		\begin{abstract}
			The analysis of single particle trajectories plays an important role in elucidating dynamics within complex environments such as those found in living cells. However, the characterization of intracellular particle motion is often confounded by confinement of the particles within non-trivial subcellular geometries. Here, we focus specifically on the case of particles undergoing Brownian motion within a tubular network, as found in some cellular organelles. An unraveling algorithm is developed to uncouple particle motion from the confining network structure, 
			 allowing for an accurate extraction of the diffusion coefficient, as well as differentiating between Brownian and fractional Langevin motion. We validate the algorithm with simulated trajectories and then highlight its application to an example system: analyzing the motion of membrane proteins confined in the tubules of the peripheral endoplasmic reticulum in mammalian cells. We show that these proteins undergo diffusive motion with a well-characterized diffusivity. 
			Our algorithm provides a generally applicable approach for disentangling geometric morphology and particle dynamics in networked architectures.

		\end{abstract}
		
	\maketitle

\section{Introduction}

Particle tracking experiments have a long history in the field of soft matter physics, where they are used to analyze the material properties of complex fluids and the dynamic behavior of active media (reviewed in \cite{wirtz2009particle,squires2010fluid}). In recent years, tracking of particles inside living cells has been extensively employed to elucidate the dynamics of cellular components ranging from single molecules~\cite{elf2007probing,parry2014bacterial,holcman2018single} to vesicular organelles~\cite{guo2014probing,gal2013particle,koslover2016disentangling,lin2016active}. Quantification of {\em in vivo} particle trajectories can be leveraged to identify state transitions and organelle interactions~\cite{lin2016active,chen2015memoryless,balint2013correlative}, to explore the rheology of intracellular fluids\cite{weihs2006bio,gal2013particle}, and to establish the underlying physical forces that drive particle motion~\cite{weber2012nonthermal,guo2014probing,fodor2015activity}.

A classic analysis approach computes the mean square displacement (MSD) of particle trajectories. In a purely thermal homogeneous system, an MSD that scales linearly with time indicates diffusive motion through a viscous fluid, with the prefactor establishing the particle diffusivity. An MSD that scales subdiffusively ($\sim t^\alpha$, $\alpha<1$) can instead indicate a viscoelastic medium with characteristic power-law scaling $\alpha$~\cite{kollmannsberger2011linear}. This analysis has been employed in a number of cellular systems, establishing diffusive behavior in the case of some protein-sized particles~\cite{wang2006single,li2015mapping,etoc2018non}, and subdiffusive dynamics for vesicles and similar-sized exogenous probes in the cytoplasm~\cite{kollmannsberger2011linear,yamada2000mechanics,hoffman2006consensus,weihs2006bio}. 
An alternate recent approach focuses instead on the velocity auto-correlation function, with negative correlations that decay as a power-law in time taken to be a sign of viscoelastic rheology~\cite{weber2012analytical,sabri2020elucidating,lucas20143d}. Still other studies analyze the distribution of individual step sizes over short time-intervals~\cite{wang2012brownian,gal2013particle,lampo2017cytoplasmic}. 

All of these approaches are confounded by a number of complications in the intracellular environment~\cite{mogre2020getting}. First and foremost, the prevalence of active forces with many different correlation timescales  implies that individual particle trajectories cannot be directly related to the rheology of the medium~\cite{guo2014probing}. Furthermore, heterogeneity in the intracellular medium indicates the existence of spatially varying diffusivities~\cite{duits2009mapping,li2015mapping} and is thought to be responsible for the broad non-Gaussian distribution of step sizes observed for many particles~\cite{wang2012brownian,lampo2017cytoplasmic,chechkin2017brownian}. Nevertheless, extracting the effectively diffusive or subdiffusive behavior of intracellular particles, albeit over a finite range of timescales, can give insight into functionally important consequences such as particle search times, kinetics, or interaction frequencies. 

In certain cases 
it is possible to disentangle the underlying particle dynamics and the effects of forces, flows, or confinement from the measured particle trajectories. For example, when particles are driven by slowly-varying fluid flows overlaid on top of diffusive behavior, the raw MSD appears super-diffusive. However, the diffusive behavior can be extracted by subtracting out a smoothed trajectory and analyzing the resulting MSD curves with an appropriate rescaling~\cite{koslover2016disentangling}. In other cases, separate measurement of the confounding factors is necessary. For instance, active microrheology measurements of cytoplasmic material properties enable `passive' particle trajectories to then be used for extraction of the spectrum of forces driving particle motion~\cite{guo2014probing}. Mapping out the underlying surface on which a particle is confined and measuring distances along that surface has been shown to enable accurate characterization of diffusing particle dynamics~\cite{adler2019conventional}. 

Confinement effects are a common source of complication in analyzing the trajectories of intracellular particles. Confinement in a finite region can result in apparently subdiffusive MSD curves, on time scales up to an order of magnitude shorter than the typical time to traverse the confining region~\cite{saxton1995single}. Similarly, velocity autocorrelation functions for particles in confinement exhibit negative peaks that increase when the velocity is measured across longer time intervals~\cite{weber2012analytical}. In this manuscript, we focus on how to decouple confinement from the underlying particle dynamics in the case of particles moving along tubular networks.

Eukaryotic cells contain a number of structures that result in confinement of particles at different length scales. The extent of the cell itself provides an upper limit for confinement. Furthermore, many particles are embedded in the membrane or interior of organelles that offer confinement on a subcellular scale. One example of interest is the reticulated mitochondrial network formed in many cell types. These networks exhibit varying connectivity that can be tuned by genetic perturbation of the proteins responsible for mitochondrial fusion and fission~\cite{viana2020mitochondrial}. 
Another organelle with a networked morphology is the endoplasmic reticulum (ER), whose functional roles include lipid distribution, calcium buffering, and protein processing, sorting, and quality control~\cite{schwarz2016endoplasmic,perkins2021intertwined}. The ER forms an interconnected system of tubules and stacked sheets, with a continuous lumen and membrane, that spans throughout the cell~\cite{westrate2015form,friedman2011er}. Away from the cell nucleus, the ER can be approximated as a tubular network with largely three-way junctions, through which secretory proteins must move in order to encounter the point-like ER exit sites~\cite{chen2013er,griffing2010networking,barlowe2016cargo}. 

Mathematical models of transport within the ER and mitochondrial networks indicate that the network morphology has the potential to alter search rates and kinetics for proteins to find each other and specific targets within the network~\cite{brown2020impact,viana2020mitochondrial,scott2021diffusive}. However, examining the interplay between network architecture and kinetics requires an assumed physical model for particle motion ({\em eg:} diffusive~\cite{brown2020impact,viana2020mitochondrial}, subdiffusive~\cite{stadler2018diffusion}, or locally persistent~\cite{holcman2018single,dora2020active}) and an accurate parameterization of that model. Thus, precise empirical quantification of particle dynamics within reticulated networks is of growing interest for elucidating the structure-function relationship of these critical cellular organelles.

Most prior studies on the movement of ER membrane and luminal proteins have focused on bulk measurements of protein spread, including FRAP analysis~\cite{dayel1999diffusion,siggia2000diffusion,nehls2000dynamics} and the spatiotemporal quantification of spreading from a locally photoactivated region~\cite{holcman2018single,konno2021endoplasmic}. However, it has recently become possible to track the movement of individual proteins through the ER tubules, enabling new observations of dynamics within the ER. For example, single particle trajectories of ER luminal proteins have been found to exhibit unexpectedly fast motion along individual tubes, followed by trapping at junctions -- an effect that has been attributed to putative luminal flows over short timescales~\cite{holcman2018single}. ER membrane protein trajectories, however, do not appear subject to these rapid motions. Nevertheless, the analysis of such trajectories is complicated by their confinement within the peripheral ER network structure.

In this work we present a novel method for analysis of diffusive particle trajectories that are confined within a spatial network, consisting of tubules connected by narrow junctions. Namely, we describe how such trajectories can be `unraveled' to establish a properly sampled underlying trajectory that describes the motion of an identical particle on an infinite line. This unraveling process allows the accurate extraction of a diffusion coefficient for the particle motion by removing the confounding effects of the network structure itself. The procedure is generally applicable to analysis of any locally one-dimensional trajectories where the confining network architecture can be separately imaged. Thus, for example, it can be applied to particles moving along a single finite-length tube, or along a spine-studded  morphology such as that found in neuronal dendrites~\cite{santamaria2006anomalous,sartori2020statistical}. 

We validate the method using simulations of particles diffusing on a network, and demonstrate that the technique can differentiate between particles undergoing Brownian versus fractional Langevin motion (a common model for subdiffusion due to viscoelastic rheology~\cite{mogre2020getting,deng2009ergodic}). The technique is then applied to the analysis of membrane protein trajectories in the ER, showing that these trajectories are indeed consistent with diffusive motion and providing an estimate of their diffusivity. 

Overall, the work presented here enables quantitative analysis of particle trajectories confined on network structures, allowing for mathematically accurate decoupling of confinement effects from the underlying particle dynamics.

\begin{figure*}
	\includegraphics[width=\textwidth]{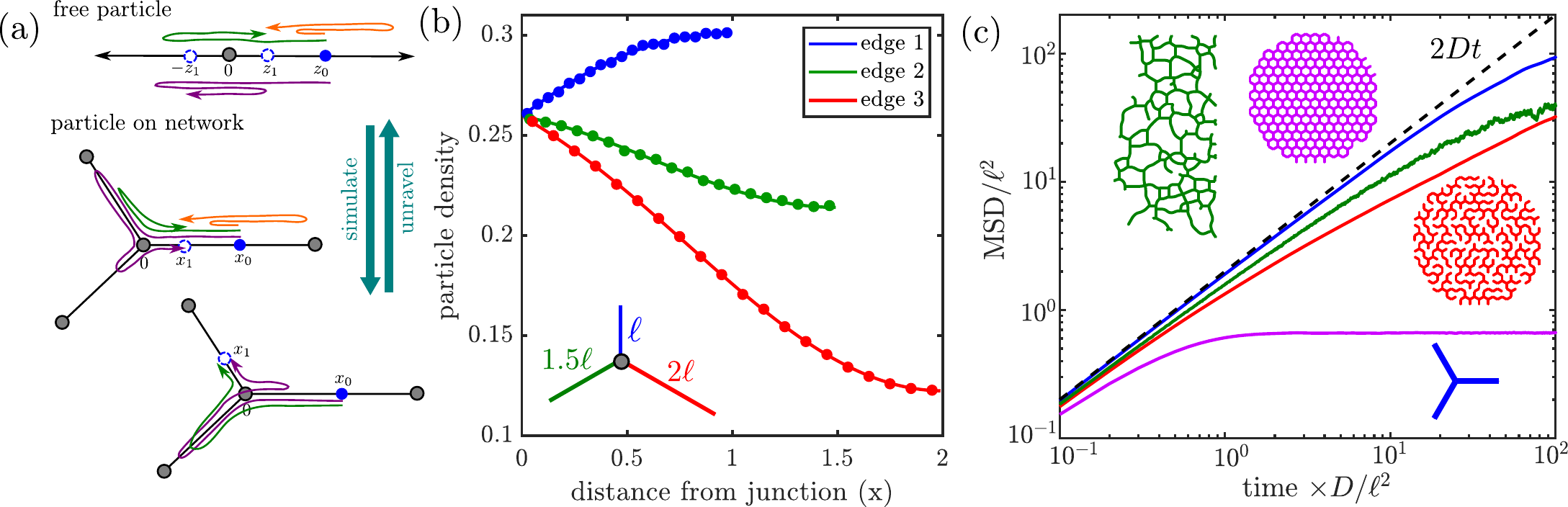}
	\caption{Simulation of diffusive particle trajectories on a network. (a) Schematic of diffusion simulation and unraveling algorithm. Top: example single-step trajectories starting at $x_0$ on an infinite line; Bottom: sample corresponding trajectories along the network. Trajectories that pass the node can land on either the same or a different edge.
		(b) Simulations of particle trajectories using the discrete time propagation algorithm (dots) reproduce exact analytic distributions (lines) on a triskelion network. Densities are shown for a particle starting at the center node, propagating for a total time $t = 0.8\ell^2/D$.
		(c) Diffusive particle mean squared displacement depends on the structure of the confining network. Dimensionless MSD is shown for particles on a triskelion geometry (blue), a honeycomb network (magenta), a honecomb with 30\% of its edges removed (red), and a section of a peripheral ER network extracted from a COS7 cell (green). Length units are scaled by the average network edge length ($\ell$) and time units by $\ell^2/D$.	 	
	}
	\label{fig:simMSD}
\end{figure*}

\section{Results}

\subsection{Simulating diffusive particles on a network}
\label{sec:simdiff}

We begin by simulating the behavior of diffusive particles on a tubular network. Individual narrow tubules are treated as effectively one-dimensional segments, joined together at point-like junctions. The network structure is thus idealized as a set of nodes connected by edges with well-defined length: $\ell_{m}$ for the $m^\text{th}$ edge. The edges are not necessarily straight, so that $\ell_m$ can be longer than the spatial distance between two adjacent nodes.

Several algorithms are possible to simulate the trajectory of a diffusive particle on such a network. In recent work, we demonstrated an exact kinetic Monte Carlo algorithm that leverages analytical propagator functions obtained by solving the diffusion equation in the neighborhood of each node~\cite{scott2021diffusive}. By sampling transition times from these functions, the particle can be propagated between adjacent nodes as a continuous-time random walk, with no discretization artifacts. Here, we take an alternate approach, which starts from a diffusive particle trajectory $z(t)$ on an infinite line and maps it onto a corresponding trajectory $x(t)$ on the network. In a subsequent section, we demonstrate how this mapping can be reversed to extract the underlying diffusive dynamics of a particle whose motion is observed on a network of known structure. 

Consider a discretized trajectory $\left\{z_i\right\}$ of a Brownian particle with diffusivity $D$, evaluated at discrete times $t_i = i \Delta t$. The increments of such a trajectory are distributed according to a normal distribution: $\Delta z \sim \mathcal{N}(0,\sqrt{2D\Delta t})$. To map this trajectory onto an equivalent particle diffusing over the network structure, we use these increments $\Delta z$ for propagation on the network. In particular, we assume that each time the diffusing particle passes a network node, it selects uniformly at random the next edge from the ones attached to that node. Our approach assumes that the discrete time step is short enough that on any single step the particle can pass only the network node that is closest to its starting position (in terms of distance along the edge).
Specifically, we assume that $|\Delta z| < \ell_m/2$ for the current edge $m$ on which the particle is located.

 To map an unconstrained trajectory step $\Delta z$  onto the network (Fig.~\ref{fig:simMSD}a), consider a particle starting at position $x_0$ along edge $m$, and assume that $x_0<\ell_m/2$, so that the particle is closer to the first node of the edge. We want to find the next position $x_1$ for the particle, given its sampled trajectory on the infinite line. If $\Delta z < -x_0$ then the particle trajectory must have passed the node during this time step. We then place the particle at position $x_1=|x_0+\Delta z|$ on an edge selected uniformly at random from among all the edges attached to that node (including the original edge).
  If $\Delta z > -x_0$ then the particle may also have passed the node during that time step. The probability that it has done so can be computed as
\begin{equation}
\begin{split}
p_\text{pass} = \exp\left[-\frac{(x_0+\Delta z)x_0}{D\Delta t}\right],
\end{split}
\label{eqn:ppass}
\end{equation}
(see Appendix~\ref{sec:simalgorithm} for derivation).
If it passed the node, then it can end the time step on any of the adjacent edges with equal probability; if it did not pass the node then it must still be on its original edge. 
The subsequent position of the particle is then given by $|x_0+\Delta z|$ with probability $1-p_\text{pass}+p_\text{pass}/d$ of landing on its original edge and probability $p_\text{pass}/d$ of landing on each of the other adjacent edges, where $d$ is the degree of the node. The propagation of the particle is then repeated from its new position. If the particle originates close to the other end of the edge ($x_0>\ell_m/2$) then an analogous calculation is performed by considering the nearby node on the far end of edge $m$. 

It should be noted that this algorithm also handles reflections off of a dead-end node. Increments that would lead an unconstrained particle to pass such a node simply result in the particle being placed back upon the single edge leading to the dead-end. Such explicit reflections are typically implemented in Brownian and fractional Brownian dynamics simulations with hard-wall boundaries\cite{jeon2010fractional}.

The proposed algorithm accurately represents the behavior of a diffusive particle over the time step $\Delta t$ so long as we can neglect any trajectories that pass a node other than the most nearby one over the course of a single time step.
Therefore, unlike the exact algorithm described in previous work~\cite{scott2021diffusive} it requires the constraint $\sqrt{2 D \Delta t} \ll \ell_m$. In Fig.~\ref{fig:simMSD}b, we demonstrate that for small $\Delta t$ the agent-based simulations described here accurately reproduce analytically computed particle distributions in a simple triskelion network (see Appendix~\ref{app:triskelion}).

\begin{figure*}
	\includegraphics[width=\textwidth]{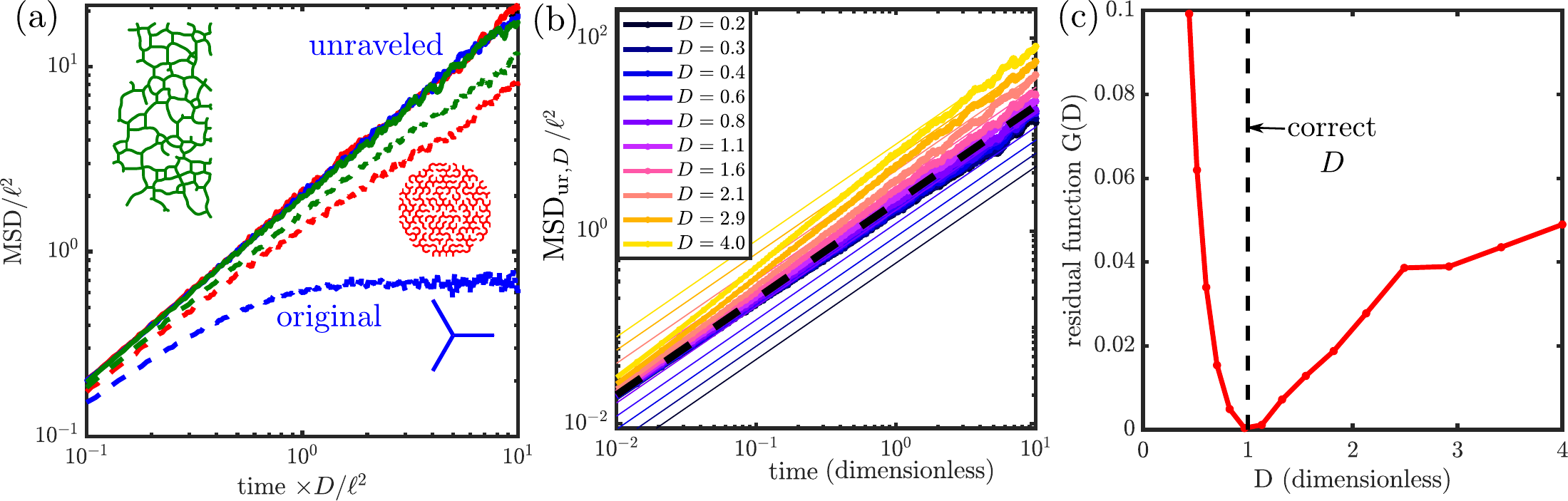}
	\caption{Unraveling simulated trajectories to estimate diffusivity. (a) Dimensionless MSD of particle trajectories on a triskelion (blue), a decimated honeycomb network (red), and a segment of peripheral ER (green). All simulations were run with $D=1$. Original MSD (from Fig.~\ref{fig:simMSD}c) is shown with dashed lines, MSD after unraveling (with $D=1$) shown with solid lines. (b) MSD of trajectories on decimated honeycomb network, unraveled using different values of the diffusivity $D$ (thick lines), compared to the expected $2Dt$ (thin lines). Dashed black line indicates values for the correct diffusivity. All units are non-dimensionalized by the edge length and the simulated diffusivity. (c) Residual function $G(D)$ plotted for the different diffusivity values used in unraveling. Dashed line shows the correct $D=1$ used in the simulation, which matches well to the minimum of the residual function.}
	\label{fig:unravelsim}
\end{figure*} 
 
\subsection{Network confinement modulates particle MSD}

A traditional metric for analyzing particle trajectories is to look at the mean squared displacement (MSD), defined by
\begin{equation}
\begin{split}
\text{MSD}(t) = \left<\left|\vec{r}(t) - \vec{r}(0)\right|^2\right>,
\end{split}
\end{equation}
where $\vec{r}(t)$ gives the spatial position of the particle at time $t$, and the average is taken both over time and over an ensemble of many particles. Particles that diffuse unimpeded on an infinite linear domain exhibit the relationship $\text{MSD} = 2Dt$. Identical behavior is observed for particles diffusing on a network comprising an infinite lattice of tubules. The effect of confinement in a finite domain is a well-known complication of MSD analysis~\cite{saxton1995single}, with the MSD beginning to flatten on time scales comparable to $\sqrt{R^2/2D}$ for a domain of radius $R$ (Fig.~\ref{fig:simMSD}c).

When the connectivity of the network structure is reduced below that of the well-connected lattice, the time dependence of the mean squared displacement flattens still further. Scaling analysis of percolation systems indicates that particles moving on the largest connected cluster of a network above the percolation phase transition should maintain effectively diffusive behavior, with an effective diffusivity that decreases according to $D_\text{eff} \sim (p-p_c)^\mu$, where $p$ is the neighbor connection probability, $p_c$ is its critical value, and the scaling exponent is $\mu \approx 1.3$ for planar systems~\cite{stauffer2018introduction}. However, for a finite network and over a finite observation time, the MSD on a lattice with missing connectivity can exhibit behavior that looks distinctly non-diffusive. In Fig.~\ref{fig:simMSD}c (red curve), we show the MSD of particles on a honeycomb network with $30\%$ of its edges removed in such a way as to maintain a single connected component. The MSD appears to show a subdiffusive (sublinear) scaling with time, as missing connectivity forces particles to take much longer paths to get from one part of the network to another~\cite{brown2020impact}.

Cellular organelle networks, such as the peripheral endoplasmic reticulum and mitochondrial network structures, also show reduced connectivity, with dead-end nodes and heterogeneous pore sizes between connected edges.
 Simulating diffusive particle trajectories on such networks indicates that the MSD is also expected to scale sublinearly with time (Fig.~\ref{fig:simMSD}c, green curve).

These simulations highlight the inherent difficulty of quantifying particle diffusivity on a network by analyzing the mean squared displacement of observed trajectories. Namely, the MSD convolves together two separate effects: the underlying dynamics of the moving particles and the morphology of the network within which they are confined. We thus proceed to develop a method for separating out these two effects and extracting particle diffusivity from their behavior on a known network structure.

\subsection{Unraveling diffusive trajectories on networks}
\label{sec:unravel}

To unravel the particle dynamics from the network architecture, we proceed by inverting the algorithm used to map the discrete trajectory of an unconfined particle ($\left\{z_i\right\}$) onto a trajectory over the network (Fig.~\ref{fig:simMSD}a). Given a set of discrete 2D (or 3D) observations  of the network-confined particle ($\left\{\vec{r}_i\right\}$), we seek to find the trajectory that particle would have taken if allowed to move over an infinite line. To begin, we re-express the spatial trajectory in terms of network coordinates $\left\{m_i,x_i\right\}$ that list the edge $m_i$ and position along that edge $x_i\in (0,\ell_i)$ of the particle at time point $t_i$. We then consider the start and end positions of the particle at each time-step, assuming that there is an underlying free one-dimensional trajectory such that $z_i, z_{i+1}$ were mapped to network positions $(m_i,x_i), (m_{i+1},x_{i+1})$ using the algorithm described in the previous section.

\begin{figure}
	\includegraphics[width=0.45\textwidth]{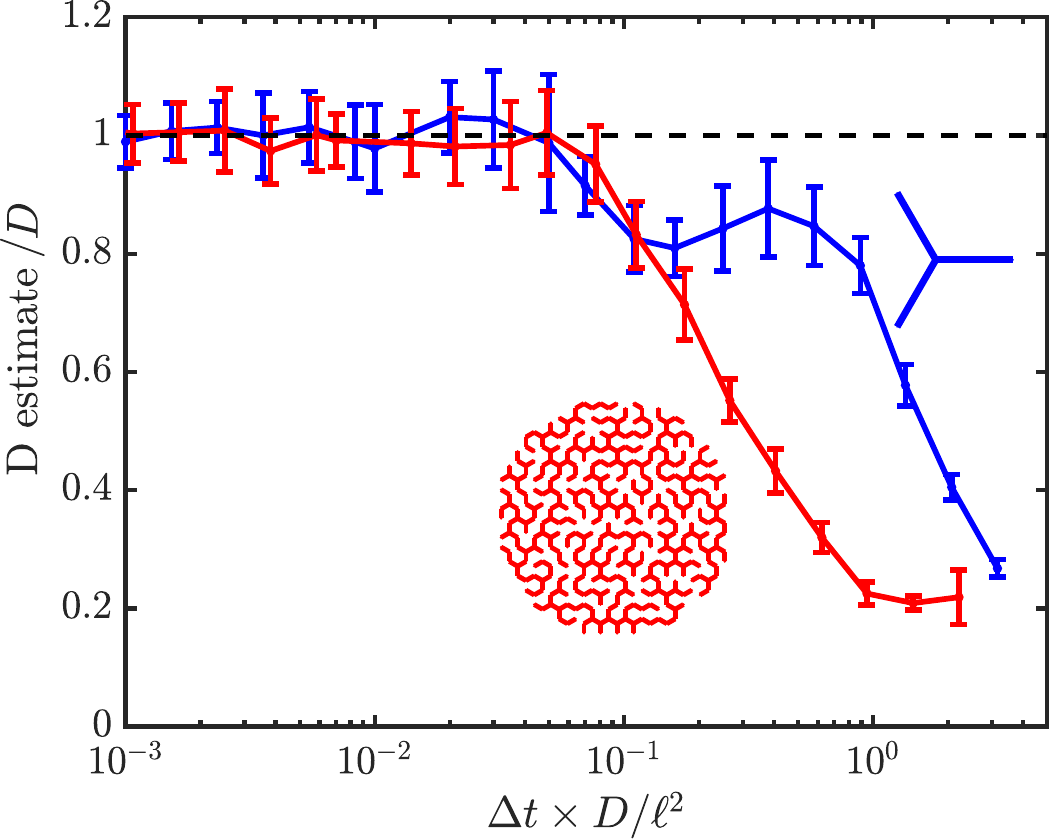}
	\caption{Diffusivity estimates for different timesteps. Simulated trajectories on a decimated honeycomb network (red) and a triskelion network (blue) are unraveled to obtain an estimated $D$ value. Each analysis involves 100 independent trajectories of 100 time steps each. Error bars give standard deviation from $20$ replicate analyses. All values are non-dimensionalized by the edge length $\ell$ and the simulated diffusivity $D$.}
	\label{fig:stepsize}
\end{figure}

For sufficiently small time-steps, we can again assume that the particle could only have passed over the node closest to $x_i$. Without loss of generality, we take this node to be the one located at position $0$ on edge $m_i$. For simplicity of notation, we shift the unconstrained trajectory such that $z_i = x_i$. The next observed position of the unconstrained particle must then be $z_{i+1} = \pm x_{i+1}$. Bayes' rule~\cite{ross2014introduction} allows calculation of the probability associated with each choice, conditional on the observed particle positions. The probability that the unconstrained trajectory ended at $z_{i+1} = x_{i+1}$ after the time-step, given that the particle ended at $(m_{i+1},x_{i+1})$ on the network is:
\begin{equation}
\begin{split}
p_+ = \mathcal{P}(z_{i+1}=x_{i+1} | m_{i+1}, x_{i+1} ) = \\
\frac{\mathcal{P}(m_{i+1} , x_{i+1}|z_{i+1}=x_{i+1})\mathcal{P}(z_{i+1}=x_{i+1})}{\mathcal{P}(m_{i+1} , x_{i+1})}
\end{split}
\label{eqn:bayes}
\end{equation}
The free particle increments are sampled from a normal distribution, so that 
\begin{equation}
\begin{split}
\mathcal{P}(z_{i+1}=\pm x_{i+1}) = \frac{1}{\sqrt{4\pi D\Delta t}} e^{-\frac{(x_i \mp x_{i+1})^2}{4Dt}}.
\end{split}
\end{equation}

If the particle moved on to a different edge ($m_{i+1} \neq m_i$), then the unconstrained trajectory must have passed zero and the conditional probabilities of the observed end-point on the network are:
\begin{eqnarray}
\mathcal{P}(m_{i+1} , x_{i+1}|z_{i+1}=x_{i+1}) & = & \frac{1}{d} e^{-\frac{x_i x_{i+1}}{D\Delta t}}, \\
\mathcal{P}(m_{i+1} , x_{i+1}|z_{i+1}=-x_{i+1}) & = & \frac{1}{d}.
\end{eqnarray}
The first line gives the probability that the unconstrained trajectory passed $0$ (as in Eq.~\ref{eqn:ppass}) before returning to the positive side. The second line simply indicates that if the unconstrained trajectory ends on the negative side, it must have passed zero and is thus equally likely to be mapped to any of the adjacent edges.

For a particle that ends on the same edge as its starting point ($m_{i+1} = m_i$), the conditional probability given $z_{i+1} = x_{i+1}$ is altered to
\begin{equation}
\begin{split}
\mathcal{P}(m_i,x_{i+1}|z_{i+1}=x_{i+1}) & = 1 + \(\frac{1}{d} - 1\) e^{-\frac{x_i x_{i+1}}{D\Delta t}}, 
\end{split}
\end{equation}
which includes an additional term for unconstrained particles that never pass the origin.

Plugging into Bayes' rule (Eq.~\ref{eqn:bayes}) results in the following probabilities for the end-point of the unconstrained trajectory:
\begin{equation}
p_+ = \begin{cases}
\frac{1}{2}, & \quad  m_{i+1} \neq m_i\\
 \frac{1}{2 -d + d\exp(x_ix_{i+1}/Dt)}, & \quad  m_{i+1} = m_i.
\end{cases}
\label{eqn:zsamp}
\end{equation}
The conditional probability $p_-$ that $z_{i+1} = -x_{i+1}$ given the observed values of $m_{i+1},x_{i+1}$ can be computed as $p_- = 1- p_+$.

For the network trajectory $\left\{m_i,x_i\right\}$, we obtain an `unraveled' trajectory $\left\{z_i\right\}$, by sampling the next position at each step according to Eq.~\ref{eqn:zsamp}, considering only the node closest to $x_i$ for potential passage by the unconstrained particle. The mean squared displacement and velocity autocorrelation function for the unraveled trajectory can then be analyzed in the usual manner.
As shown in Fig.~\ref{fig:unravelsim}a, the unraveled trajectory mean squared displacement ($\text{MSD}_\text{ur}$) for simulated particles diffusing on a network regains its linear time scaling, with the correct prefactor: $\text{MSD}_\text{ur} = 2Dt$.

Computing the unraveled trajectories requires knowledge of the particle diffusivity $D$ in order to correctly sample the step direction from Eq.~\ref{eqn:zsamp}. Since one of the primary goals of the unraveling procedure is to gain an estimate of particle diffusivity, $D$ cannot be assumed {\em a priori}. Instead, the unraveling process is performed with a range of different diffusivities to find a self-consistent value. Values of $D$ that are too large or too small result in an unraveled MSD that deviates from the expected behavior (Fig.~\ref{fig:unravelsim}b). We estimate the underlying particle diffusivity by finding the value of $D$ that minimizes the difference between the computed MSD$_\text{ur}(t;D)$ and $2Dt$. Specifically, we compute normalized squared residuals on log-log axes to assess the goodness of fit:
\begin{equation}
\begin{split}
G(D) =  \frac{\sum_i \log^2\left[\text{MSD}_\text{ur}(t_i|D)/(2Dt_i)\right]}{\sum_i \log^2\left[\text{MSD}_\text{ur}(t_i|D)/\left<\text{MSD}_\text{ur}\right>_t(D)\right]}
\end{split}
\label{eq:residuals}
\end{equation}
where the residuals are evaluated at logarithmically spaced time points $t_i$ and $\left<\text{MSD}_\text{ur}\right>_t$ refers to the average value over all time-points. The final diffusivity estimate is the value of $D$ that minimizes $G(D)$ (Fig.~\ref{fig:unravelsim}c). We note that the choice of goodness-of-fit function is equivalent to $1-R^2$ for the standard $R^2$ metric used to assess the accuracy of a linear regression, following a logarithmic transform.

Figure~\ref{fig:stepsize} shows how the estimated diffusivity varies with the time-step of the trajectory. Our approach is valid only if the time-step is small enough that individual consecutive snapshots are very unlikely to involve the particle trajectory passing any network node other than the one nearest to the particle. During the analysis, any steps that involve the particle jumping onto a non-adjacent edge result in breaking the trajectory up into separate segments. For larger step-sizes (roughly, $\Delta t \gtrsim 0.1 \ell^2/D$), such large steps become increasingly common. Breaking the trajectory then results in a systematic underestimate of the diffusivity since large steps are removed from the analysis. Thus, in order to accurately assess particle dynamics on a network, the particle positions must be visualized rapidly enough that the particle passes no more than one node on each step.

\subsection{Fractional Langevin motion and velocity autocorrelations}

A common application of MSD analysis for single particle trajectories is to elucidate the relevant dynamic model that best describes their behavior\cite{gal2013particle}. In particular, particles that exhibit an MSD scaling linearly with time are generally assumed to be moving in an effectively diffusive manner. By contrast, subdiffusive scaling ($\text{MSD} \sim t^\alpha$ for $0<\alpha < 1$) can be explained by a number of underlying physical models, including continuous time random walks, confinement, and fractional Langevin motion (fLM)~\cite{weber2012analytical,metzler2000random}. The latter model has been shown to apply to a variety of cytoplasmic and intranuclear particles~\cite{lucas20143d,lampo2017cytoplasmic,weber2010bacterial,sabri2020elucidating}. Fractional Langevin motion is expected for particles subject to thermal fluctuations in a viscoelastic medium, driven by fractional Brownian forces with a power-law correlation over time that scales as $t^{-\alpha}$\cite{jeon2010fractional,bressloff2013stochastic}. 
We set out to establish whether unraveled trajectories can be leveraged to distinguish if particles are undergoing classical Brownian motion or fLM, decoupling the underlying particle dynamics from the confinement effect of the network.

To simulate the dynamics of a particle undergoing fLM on a network, we first generate fractional Brownian forces in two dimensions (2D) using previously established methods~\cite{deng2009ergodic,kroese2013spatial}. Particle step increments in 2D are then computed by convolving the past history of steps with the appropriate power-law memory kernel, followed by zeroing out of the step component perpendicular to the current edge axis (details in Appendix~\ref{app:fLM}). The overall approach is analogous to that used for simulating the fractional Langevin equation in a confined geometry~\cite{jeon2010fractional}, in the limit of no inertia and narrow edge confinement.

\begin{figure}
	\centerline{\includegraphics[width=0.48\textwidth]{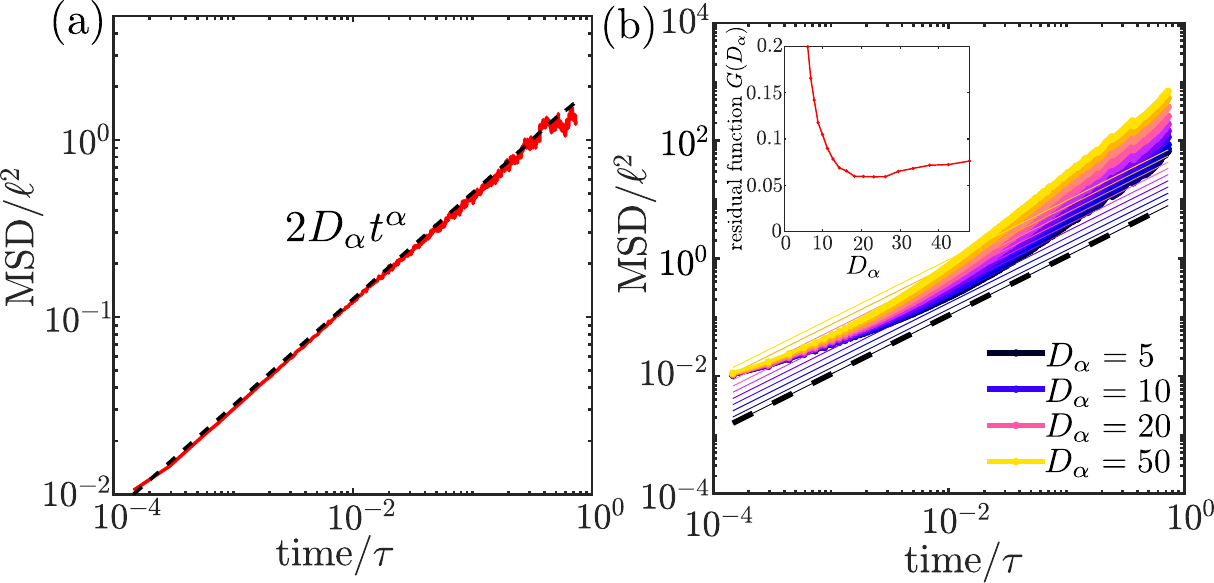}}
	\caption{Unraveling of simulated fractional Langevin motion trajectories on a decimated honeycomb network.  (a) MSD of raw trajectories, with dashed line showing expected scaling. (b) MSD of trajectories unraveled with different input values of $D_\alpha$, as in Fig.~\ref{fig:unravelsim}b. Inset shows residual function $G(D_\alpha)$ for the unraveling attempts with different $D$. Simulations used scaling exponent $\alpha=0.6$, with prefactor set such that individual step sizes are comparable to the Brownian simulations in Fig.~\ref{fig:unravelsim}, with $2D_\alpha \Delta t^\alpha = (0.1\ell)^2$. Length and time units are non-dimensionalized by $\ell$ and $\tau = (\ell^2/D_\alpha)^{1/\alpha}$, respectively.}
	\label{fig:simfbm}
\end{figure}

\begin{figure*}
	\centerline{\includegraphics[width=\textwidth]{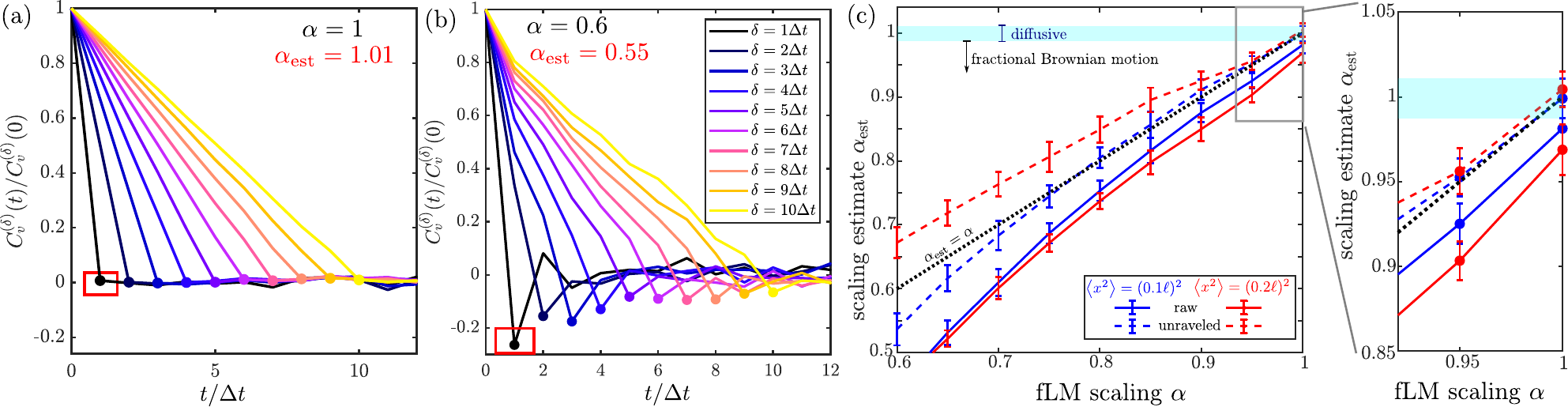}}
	\caption{Estimating scaling exponent $\alpha$ from velocity autocorrelation function of unraveled trajectories. Normalized velocity autocorrelations are shown for unraveled trajectories of simulated particles undergoing (a) Brownian motion and (b) fractional Langevin motion with $\alpha=0.6$. Red box indicates first negative peak, used to extract estimated scaling exponent $\alpha_\text{est}$. The individual step size is set such that $\left<\Delta x^2\right> = 2D_\alpha \Delta t^\alpha = (0.1\ell)^2$. %Time units are non-dimensionalized by $\tau = (\ell^2/D_\alpha)^{1/\alpha}$, respectively.
		 (c) Estimated scaling exponent for simulated fLM particles with different true values of $\alpha$. Estimations are shown based on the velocity autocorrelation of either raw 2D trajectories (solid) or unraveled 1D trajectories (dashed). Simulations were carried out with two different root mean squared step sizes: $10\%$ (blue) and $20\%$ (red) of the network edge length. 
		All simulations included $100$ particles tracked for $100$ time steps. Error bars are standard deviations from $50$ independent replicates. Cyan bar indicates the range of $\alpha_\text{est}$ that should be taken to imply purely diffusive motion, obtained as average $\pm$ standard deviation for Brownian particle simulations. Right: zoomed-in section shows that network confinement makes Brownian particles ($\alpha=1$) appear to have a lowered estimated scaling ($\alpha_\text{est}<1$), whereas unraveled trajectories show the correct Brownian exponent $\alpha_\text{est}=1$.}
	\label{fig:velautocorr}
\end{figure*}

As expected, the simulated trajectories exhibit an MSD that scales as a power-law with time, $\text{MSD} \sim t^\alpha$  (Fig.~\ref{fig:simfbm}a). These trajectories can be unraveled as described in the previous section, to remove the effect of network confinement. However, because the unraveling procedure assumes diffusive transport when computing node passage probabilities, the resulting unraveled MSD curves are notably deformed (Fig.~\ref{fig:simfbm}b). Consequently, the goodness-of-fit metric (comparing to a linear, diffusive scaling) is substantially higher for these trajectories, serving as an indicator of non-diffusive transport.

An alternative approach for distinguishing fLM from diffusive trajectories on a network involves analysis of the velocity autocorrelation function for the unraveled trajectories. For velocities computed over time-interval $\delta$, the autocorrelation function is defined by~\cite{weber2010subdiffusive}
\begin{equation}
\begin{split}
C^{(\delta)}_v (t) = \frac{1}{\delta^2} \left< [x(t+\delta)-x(t)] \cdot [x(\delta)-x(0)] \right>,
\end{split}
\end{equation}
where the average is taken over both time and the particle ensemble. This function exhibits a negative peak at $t=\delta$, rising back towards zero in a polynomial fashion for $t>\delta$. The normalized negative peak $C_v^{(\delta)}(\delta)/C_v^{(\delta)}(0)$ becomes deeper for more subdiffusive motion (lower $\alpha$) and, for fractional Langevin motion, is independent of the choice of window size $\delta$ in defining the velocity\cite{sabri2020elucidating,lucas20143d,weber2012analytical}. 
The scaling exponent $\alpha$ can be extracted directly from this value according to the expression~\cite{weber2010subdiffusive}:
\begin{equation}
\begin{split}
\left|\frac{C_v^{(\delta)}(\delta)}{C_v^{(\delta)}(0)}\right| = 1 - 2^{\alpha-1}.
\end{split}
\label{eq:negpeak}
\end{equation}

We plot the normalized velocity autocorrelation for unraveled trajectories of Brownian particles and particles undergoing fLM in Fig.~\ref{fig:velautocorr}a,b. For Brownian particles, the velocity correlations are flat for $t>\delta$, emphasizing that the unraveling procedure removes the effect of confinement and dead-ends in the network, which would be expected to yield negative peaks~\cite{weber2012analytical}. For particles undergoing fLM, the unraveling decreases the magnitude of the negative peaks at larger step-sizes, consistent with the turning upward of the MSD curves for these trajectories. However, the form of the velocity autocorrelation function is still clearly non-diffusive, demonstrating that this metric can be used to diagnose fLM versus Brownian motion on a network.

\begin{figure*}
	\includegraphics[width=\textwidth]{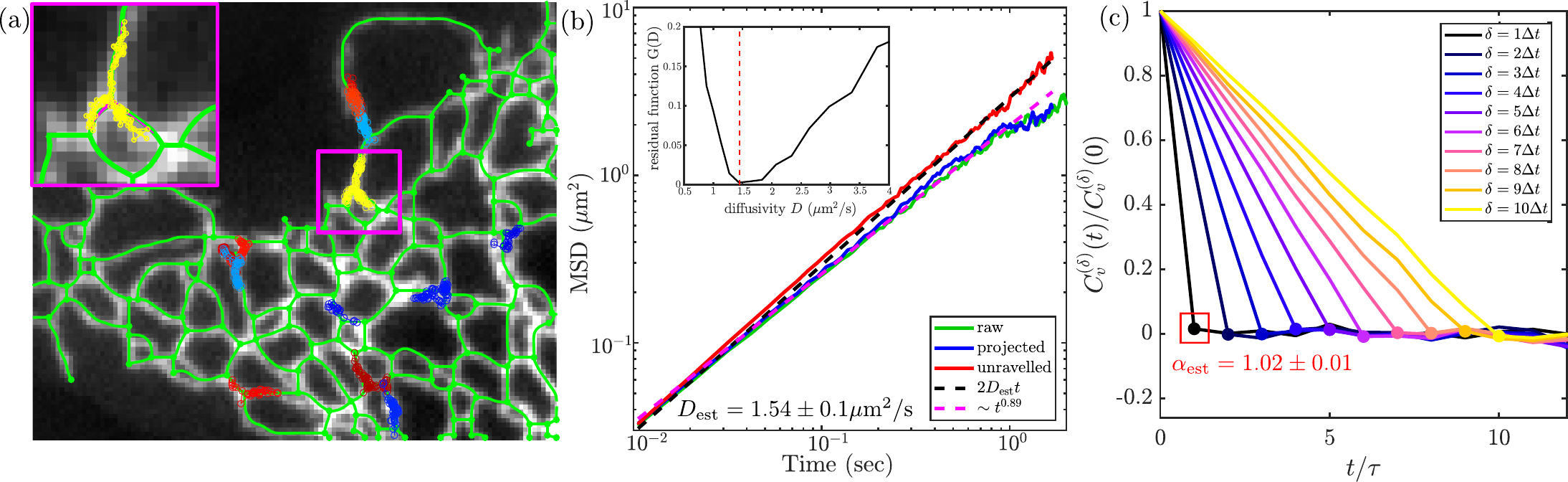}
	\caption{Unraveling of membrane protein (Halo-TA) trajectories in the peripheral ER networks of an example COS7 cell. (a) Extracted network structure (green) superimposed on a confocal image with fluorescently labeled ER marker. Trajectories within $1$ sec of this snapshot are shown as colored circles. Inset shows single trajectory, projected onto network edges (magenta). (b) Ensemble-averaged MSD for raw trajectories (green), projected trajectories (blue), and unraveled trajectories (red). Dashed magenta line indicates apparent subdiffusion for particle trajectories due to confinement in network structure. Dashed black line shows linear scaling with the estimated diffusivity. Inset: residual function for unraveling with different diffusivity values, with self-consistent estimate marked in red. (c) Normalized velocity autocorrelation for unraveled trajectories, with estimated scaling exponent $\alpha_\text{est}$ is obtained from the first negative peak. Errors for $D_\text{est}$ and $\alpha_\text{est}$ are standard deviations obtained from bootstrapping over individual unraveled trajectories.}
	\label{fig:expttraj}
\end{figure*}

To generate an estimate of the fLM scaling exponent, we use Eq.~\ref{eq:negpeak} at $\delta = \Delta t$ (a single trajectory step).
As shown in Fig.~\ref{fig:velautocorr}c, unraveled trajectories of simulated Brownian particles on a network give an accurate estimate of $\alpha_\text{est} \approx 1.0 \pm 0.01$ (mean $\pm$ standard deviation) based on the first negative peak in the velocity autocorrelation functions. When the same analysis is applied to simulated fLM trajectories with $\alpha=0.6$, the unraveling procedure gives a somewhat biased estimate of the scaling exponent ($\alpha_\text{est} = 0.54 \pm 0.025$), but still clearly differentiates between fLM and Brownian motion. Simulations of fLM with a range of different scaling exponents (Fig.~\ref{fig:velautocorr}c) indicate that accurate estimates of the scaling exponent can be obtained for $\alpha\gtrsim 0.7$, when individual step sizes are set to $10\%$ of the network edge length. For larger step sizes, this approach overestimates $\alpha$, due to the bias towards diffusive motion introduced in the unraveling process whenever the particle passes a node. However, particles undergoing fLM still yield an estimated $\alpha_\text{est}$ that is significantly less than $1$, allowing the trajectories to be distinguished from those of Brownian particles. By contrast, if the velocity autocorrelation is analyzed for raw trajectories confined on a network, then even Brownian particles ($\alpha=1$) are likely to yield an estimated scaling value $\alpha_\text{est}$ that is below the expected range (shaded region in Fig.~\ref{fig:velautocorr}c). Thus, network confinement can make Brownian motion appear as fLM, and the unraveling procedure removes this confounding effect.

\subsection{Application to membrane proteins on ER tubules}

As a direct application of this methodology to experimental data, we consider single particle trajectories of a model ER transmembrane protein (a HaloTag fused to the minimal targeting domain of Sec61b, Halo-TA), moving within the peripheral ER tubules of COS7 cells. Widefield images of the ER network as well as individual fluorescently labeled Halo-TA proteins were collected simultaneously with rapid time resolution ($100$Hz). Because ER networks are dynamic, rearranging over tens of seconds~\cite{friedman2010er,lin2017modeling,lee1988dynamic,lee1989construction}, peripheral network structures were extracted at $1$ second intervals to serve as the underlying domain geometry for the particle trajectories. Details on imaging, image analysis, and particle tracking methods are provided in the Methods section (Appendix ~\ref{app:methods}).

For each network structure, trajectory segments collected within $\pm 1$ sec of the corresponding image time were mapped onto the network by projecting to the nearest point along the edges (Fig.~\ref{fig:expttraj}a). All projected trajectories imaged within a given cell (corresponding to approximately $1$ minute total imaging time) were then collected together for analysis.
As seen in Fig.~\ref{fig:unravelsim}b, both the raw and projected trajectories exhibited an apparently subdiffusive MSD scaling with time, as expected for simulated diffusive particles on a sample ER network (Fig.~\ref{fig:simMSD}c). The close correspondence between raw and projected trajectories indicates that the proteins were in fact largely confined to the visible ER tubules, with most of their motion occuring along the tubule axis. In Appendix ~\ref{app:circum} we explore further the effect of particles moving along the circumference of the ER tubules rather than directly along the axis. We show, using simulations, that the narrow 50 nm radius of ER tubules~\cite{schroeder2019dynamic} implies that these perpendicular particle displacements will not significantly affect the analysis.

Projected trajectories were unraveled as described in Section~\ref{sec:unravel} to extract the underlying free-particle dynamics. The mean squared displacement $\text{MSD}_\text{ur}$ of the unraveled trajectory closely matched the expected $2D_\text{est}t$ behavior (Fig.~\ref{fig:expttraj}b) with a diffusivity of $D_\text{est}=1.54\pm0.1 \mu\text{m}^2/\text{s}$, where the standard deviation in the estimate is obtained by bootstrapping over individual particle trajectories. The velocity autocorrelation function of the unraveled trajectories is consistent with the expected behavior for diffusive particles, showing very little correlation for $t>\delta$ (Fig.~\ref{fig:expttraj}c). Estimating the scaling exponent $\alpha$ from these velocity autocorrelations (as described for simulated trajectories above) yields a value of $\alpha_\text{est} = 1.02 \pm 0.01$, with standard deviation again obtained by bootstrapping.

\begin{figure}
	\centerline{\includegraphics[width=0.4\textwidth]{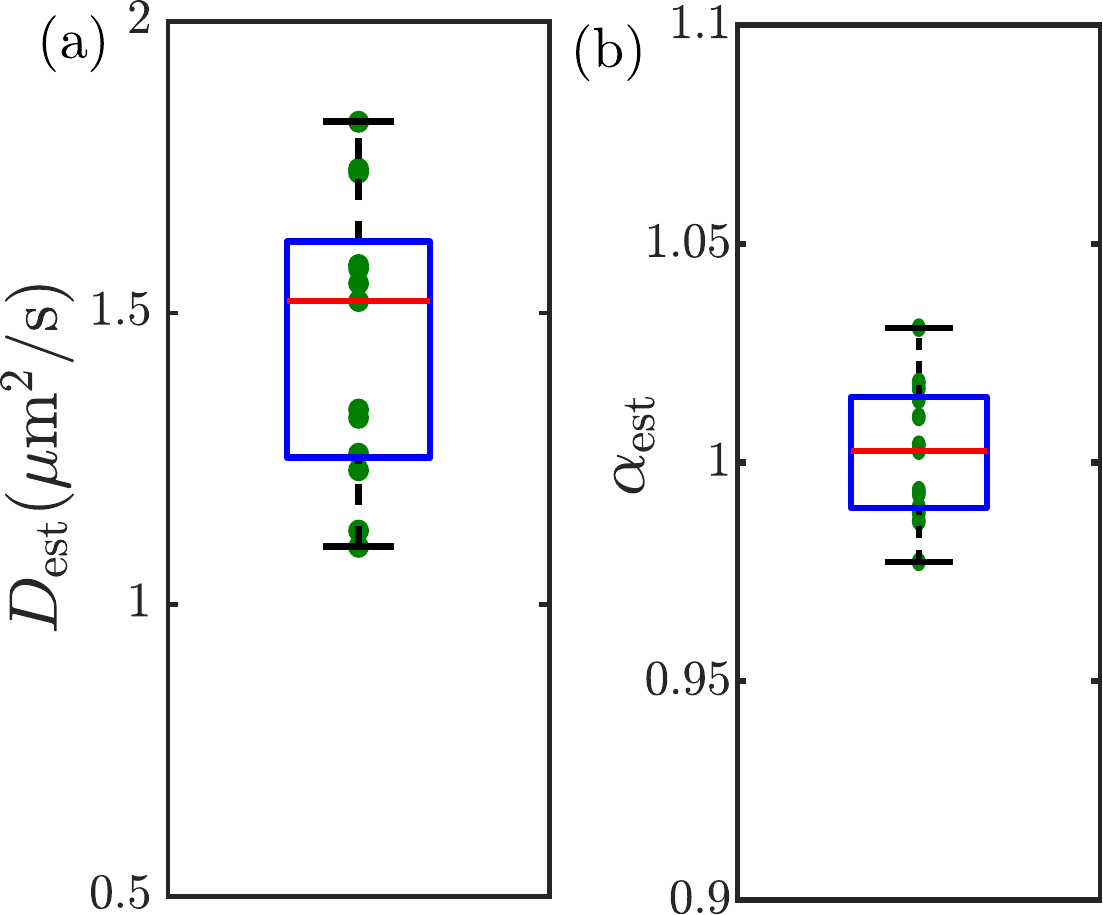}}
	\caption{Analysis of trajectories for Halo-TA ER membrane proteins from multiple COS7 cells. (a) Estimated diffusivity $D_\text{est}$. (b) Estimated scaling exponent $\alpha_\text{est}$. $N=13$ independent cells were analyzed.}
	\label{fig:allexptest}
\end{figure}

The analysis was repeated for particle trajectories and network structures obtained from $11$ different cells (Fig.~\ref{fig:allexptest}), with resulting diffusivity estimates $\left<D\right> = 1.45 \pm 0.24\mu\text{m}^2/\text{s}$ (mean $\pm$ standard deviation). The extracted diffusivity is similar to values recently estimated from bulk measurements of photoactivated protein spreading in the ER~\cite{konno2021endoplasmic}. The scaling estimates from the velocity autocorrelation of unravelled trajectories in each cell yielded $\left<\alpha\right> = 1.00 \pm 0.02$ (standard deviation across cells), consistent with diffusive motion of the Halo-TA ER membrane protein.

We note that the average edge length of the ER networks extracted in this study was $1.0\pm 0.7\mu$m (mean $\pm$ standard deviation). Given our diffusivity estimate, we would expect the unraveling procedure to be accurate for step sizes $\Delta t \lesssim 0.07\text{sec}$, a condition which is satisfied by the $100$Hz imaging rate.

\section{Discussion}

Traditional approaches to quantifying single particle trajectories involve using the mean squared displacement or velocity autocorrelation function to characterize the diffusive or subdiffusive motion of the particles. However, for particles trapped inside the complex spatial geometry of intracellular structures, these approaches are confounded by the effects of confinement. Such effects can make diffusive motion appear subdiffusive and lead to underestimation of the particle motility. 
In this study, we demonstrate a methodology for decoupling diffusive particle behavior from the effect of spatial confinement using direct observation of the structure, in the specific case of particles confined within a tubular network.

 Our approach involves unraveling a particle trajectory on a network by sampling the hypothetical path of that particle if it were to undergo unconstrained diffusion on an infinite line. 
In classic simulations of a Brownian particle, each step of the particle represents an ensemble of possible paths leading from the starting position to the next one. When all likely paths  stay within a single edge of the network, such a classically sampled step remains accurate.
However, if there is a non-trivial chance of passing a network junction during that time-step, then it becomes necessary to consider the excursions of the particle to different edges surrounding that junction. We develop an algorithm that leverages the statistical weighting of these excursions to `fold' one-dimensional diffusive paths onto the edges around a network junction. This serves as a mathematically accurate method for discrete-time Brownian dynamics simulations of particles along a network of tubules.

We then take a Bayesian approach to inverting this algorithm. Namely, starting with an observed trajectory of discrete steps on a network, we sample the unconfined one-dimensional trajectory that would lead to the observed particle positions. The resulting `unraveled' trajectories of simulated Brownian particles exhibit all the features of classic Brownian motion and can be analyzed using the usual quantification of MSD and velocity autocorrelations.
This approach makes it possible to extract the diffusivity of the particle as well as to differentiate between diffusive and subdiffusive motion.

Simulations of particles undergoing fractional Langevin motion on network structures indicate that applying our unraveling algorithm can accurately differentiate such dynamics from Brownian diffusion. For sufficiently small step-sizes, analysis of the velocity autocorrelation function for unraveled trajectories yields an accurate estimate of the fractional scaling exponent $\alpha$.
  Furthermore, although raw Brownian trajectories may appear subdiffusive due to confinement within the network, unraveling the trajectories allows for accurate estimates of Brownian scaling behavior ($\alpha=1$) from the velocity autocorrelation function. Hence, the unraveling technique can be used both to diagnose particle motion as diffusive (as opposed to fLM) and to estimate the associated diffusivity, while removing the confounding effects of network confinement.

Our proposed approach for uncoupling diffusive transport from network confinement is subject to certain inherent limitations. It is applicable specifically to particles moving on a network of narrow tubules, so that motion along individual edges can be approximated as one-dimensional. It also requires sufficiently frequent observations of the particle position so that each particle can be assumed to pass at most one node junction during each time-step. This is equivalent to requiring that the particle steps be shorter than the typical edge length. Expanding the proposed algorithm to find the correct statistical weighting for paths that pass multiple nodes over the course of a time-step serves as a promising avenue for future work.

As an example application of the unraveling algorithm, we analyze the trajectories of membrane-bound proteins in the endoplasmic reticulum (ER). The peripheral ER in many cell types forms a tubular network structure, which hosts a variety of proteins that diffuse within its lumen and membrane. Among its many biologically crucial functions, it is thought to serve as an intracellular  network for the sorting and delivery of ions, lipids, and secreted proteins. Currently, the mechanism and spatial dynamics of these phenomena are challenging to observe, since the complex structure masks the interpretation of dynamic data collected within the organelle. Quantifying particle motion in the ER is crucial to elucidating the impact of its reticulated structure on the kinetics of biochemical pathways embedded within this organelle. Such studies constitute a first step towards generating a mechanistic understanding of ER-localized biological processes and their regulation. 

Trajectories of single transmembrane proteins on mammalian ER networks were collected at a sampling rate of 100Hz, sufficiently rapid to ensure multiple steps of each protein between junction passage events. Traditional analysis of the MSD for raw particle trajectories was confounded by confinement within the network, as indicated by an apparent subdiffusive scaling estimate. However, 
 application of the unraveling algorithm demonstrated that the dynamics of these particles is consistent with classic Brownian motion, with a diffusivity of $D\approx 1.5\mu\text{m}^2/\text{s}$. This estimate is compatible with bulk measurements of protein diffusivity in the ER membrane~\cite{konno2021endoplasmic}.

The algorithm proposed here is not specific to protein motion in the ER, but rather is a general mathematical tool for analyzing ensembles of diffusive trajectories on networks embedded in physical space. Such spatial networks have been analyzed in the context of roads, power grids, and venation patterns as well as intracellular structures~\cite{barthelemy2011spatial}. In cell biology, mitochondrial networks~\cite{viana2020mitochondrial} and the dendritic trees of neurons~\cite{sartori2020statistical} provide additional examples of network structures where diffusion can play an important role in protein transport. Our focus here is specifically on diffusive dynamics over the network, but the unraveling method also provides a potential tool for diagnosing when particle motion is non-diffusive. We show that it can accurately distinguish fractional Langevin motion from true diffusion. Future work will pursue the possibility of distinguishing other movement patterns, such as directed flow along edges~\cite{holcman2018single}, or continuous-time random walks associated with sporadic binding~\cite{malchus2010anomalous}. 

Because the unraveling approach provides a quantitative estimate of particle diffusivity, it has potential applications in distinguishing the effect of external perturbations on either network structure or the dynamics of the particles themselves. For example, mutations in ER morphogen proteins can alter the radius of ER tubules~\cite{konno2021endoplasmic}, or the density of tubule junctions~\cite{wang2016cooperation}. However, the consequences of these morphological changes on protein diffusivity in the ER remain unclear, and could be approached by unraveling membrane and luminal protein trajectories according to the methods proposed here. Overall, the ability to quantify trajectories by removing the complicating effects of confinement  is critical to understanding the behavior of particles trapped in complex reticulated geometries, in cell biology and beyond.

\section*{Acknowledgements}
This work was initiated at the QCB Cell Modeling Hackathon, supported by Wallace Marshall and the Quantitative Cell Biology Network (sponsored by NSF grant MCB‐1411898). The authors thank Jonathon Nixon-Abell, Zubenelgenubi Scott, and the members of the Hackathon for helpful discussions. They also thank Jonathon Nixon-Abell for collecting and curating some of the data used in this study. Funding for this work was provided by NSF grant MCB-2034482 (EFK), a Cottrell Scholar Award from the Research Corporation for Science Advancement (EFK), a Chancellor Funded Research Grant (EFK), and direct support of the Howard Hughes Medical Institute (CJO and JL-S).

\bibliography{networkTraj}

%merlin.mbs apsrev4-1.bst 2010-07-25 4.21a (PWD, AO, DPC) hacked
%Control: key (0)
%Control: author (8) initials jnrlst
%Control: editor formatted (1) identically to author
%Control: production of article title (-1) disabled
%Control: page (0) single
%Control: year (1) truncated
%Control: production of eprint (0) enabled
\begin{thebibliography}{72}%
\makeatletter
\providecommand \@ifxundefined [1]{%
 \@ifx{#1\undefined}
}%
\providecommand \@ifnum [1]{%
 \ifnum #1\expandafter \@firstoftwo
 \else \expandafter \@secondoftwo
 \fi
}%
\providecommand \@ifx [1]{%
 \ifx #1\expandafter \@firstoftwo
 \else \expandafter \@secondoftwo
 \fi
}%
\providecommand \natexlab [1]{#1}%
\providecommand \enquote  [1]{``#1''}%
\providecommand \bibnamefont  [1]{#1}%
\providecommand \bibfnamefont [1]{#1}%
\providecommand \citenamefont [1]{#1}%
\providecommand \href@noop [0]{\@secondoftwo}%
\providecommand \href [0]{\begingroup \@sanitize@url \@href}%
\providecommand \@href[1]{\@@startlink{#1}\@@href}%
\providecommand \@@href[1]{\endgroup#1\@@endlink}%
\providecommand \@sanitize@url [0]{\catcode `\\12\catcode `\$12\catcode
  `\&12\catcode `\#12\catcode `\^12\catcode `\_12\catcode `\%12\relax}%
\providecommand \@@startlink[1]{}%
\providecommand \@@endlink[0]{}%
\providecommand \url  [0]{\begingroup\@sanitize@url \@url }%
\providecommand \@url [1]{\endgroup\@href {#1}{\urlprefix }}%
\providecommand \urlprefix  [0]{URL }%
\providecommand \Eprint [0]{\href }%
\providecommand \doibase [0]{http://dx.doi.org/}%
\providecommand \selectlanguage [0]{\@gobble}%
\providecommand \bibinfo  [0]{\@secondoftwo}%
\providecommand \bibfield  [0]{\@secondoftwo}%
\providecommand \translation [1]{[#1]}%
\providecommand \BibitemOpen [0]{}%
\providecommand \bibitemStop [0]{}%
\providecommand \bibitemNoStop [0]{.\EOS\space}%
\providecommand \EOS [0]{\spacefactor3000\relax}%
\providecommand \BibitemShut  [1]{\csname bibitem#1\endcsname}%
\let\auto@bib@innerbib\@empty
%</preamble>
\bibitem [{\citenamefont {Wirtz}(2009)}]{wirtz2009particle}%
  \BibitemOpen
  \bibfield  {author} {\bibinfo {author} {\bibfnamefont {D.}~\bibnamefont
  {Wirtz}},\ }\href@noop {} {\bibfield  {journal} {\bibinfo  {journal} {Ann Rev
  Biophys}\ }\textbf {\bibinfo {volume} {38}},\ \bibinfo {pages} {301}
  (\bibinfo {year} {2009})}\BibitemShut {NoStop}%
\bibitem [{\citenamefont {Squires}\ and\ \citenamefont
  {Mason}(2010)}]{squires2010fluid}%
  \BibitemOpen
  \bibfield  {author} {\bibinfo {author} {\bibfnamefont {T.~M.}\ \bibnamefont
  {Squires}}\ and\ \bibinfo {author} {\bibfnamefont {T.~G.}\ \bibnamefont
  {Mason}},\ }\href@noop {} {\bibfield  {journal} {\bibinfo  {journal} {Annu
  Rev Fluid Mech}\ }\textbf {\bibinfo {volume} {42}},\ \bibinfo {pages} {413}
  (\bibinfo {year} {2010})}\BibitemShut {NoStop}%
\bibitem [{\citenamefont {Elf}\ \emph {et~al.}(2007)\citenamefont {Elf},
  \citenamefont {Li},\ and\ \citenamefont {Xie}}]{elf2007probing}%
  \BibitemOpen
  \bibfield  {author} {\bibinfo {author} {\bibfnamefont {J.}~\bibnamefont
  {Elf}}, \bibinfo {author} {\bibfnamefont {G.-W.}\ \bibnamefont {Li}}, \ and\
  \bibinfo {author} {\bibfnamefont {X.~S.}\ \bibnamefont {Xie}},\ }\href@noop
  {} {\bibfield  {journal} {\bibinfo  {journal} {Science}\ }\textbf {\bibinfo
  {volume} {316}},\ \bibinfo {pages} {1191} (\bibinfo {year}
  {2007})}\BibitemShut {NoStop}%
\bibitem [{\citenamefont {Parry}\ \emph {et~al.}(2014)\citenamefont {Parry},
  \citenamefont {Surovtsev}, \citenamefont {Cabeen}, \citenamefont {O’Hern},
  \citenamefont {Dufresne},\ and\ \citenamefont
  {Jacobs-Wagner}}]{parry2014bacterial}%
  \BibitemOpen
  \bibfield  {author} {\bibinfo {author} {\bibfnamefont {B.~R.}\ \bibnamefont
  {Parry}}, \bibinfo {author} {\bibfnamefont {I.~V.}\ \bibnamefont
  {Surovtsev}}, \bibinfo {author} {\bibfnamefont {M.~T.}\ \bibnamefont
  {Cabeen}}, \bibinfo {author} {\bibfnamefont {C.~S.}\ \bibnamefont
  {O’Hern}}, \bibinfo {author} {\bibfnamefont {E.~R.}\ \bibnamefont
  {Dufresne}}, \ and\ \bibinfo {author} {\bibfnamefont {C.}~\bibnamefont
  {Jacobs-Wagner}},\ }\href@noop {} {\bibfield  {journal} {\bibinfo  {journal}
  {Cell}\ }\textbf {\bibinfo {volume} {156}},\ \bibinfo {pages} {183} (\bibinfo
  {year} {2014})}\BibitemShut {NoStop}%
\bibitem [{\citenamefont {Holcman}\ \emph {et~al.}(2018)\citenamefont
  {Holcman}, \citenamefont {Parutto}, \citenamefont {Chambers}, \citenamefont
  {Fantham}, \citenamefont {Young}, \citenamefont {Marciniak}, \citenamefont
  {Kaminski}, \citenamefont {Ron},\ and\ \citenamefont
  {Avezov}}]{holcman2018single}%
  \BibitemOpen
  \bibfield  {author} {\bibinfo {author} {\bibfnamefont {D.}~\bibnamefont
  {Holcman}}, \bibinfo {author} {\bibfnamefont {P.}~\bibnamefont {Parutto}},
  \bibinfo {author} {\bibfnamefont {J.~E.}\ \bibnamefont {Chambers}}, \bibinfo
  {author} {\bibfnamefont {M.}~\bibnamefont {Fantham}}, \bibinfo {author}
  {\bibfnamefont {L.~J.}\ \bibnamefont {Young}}, \bibinfo {author}
  {\bibfnamefont {S.~J.}\ \bibnamefont {Marciniak}}, \bibinfo {author}
  {\bibfnamefont {C.~F.}\ \bibnamefont {Kaminski}}, \bibinfo {author}
  {\bibfnamefont {D.}~\bibnamefont {Ron}}, \ and\ \bibinfo {author}
  {\bibfnamefont {E.}~\bibnamefont {Avezov}},\ }\href@noop {} {\bibfield
  {journal} {\bibinfo  {journal} {Nat Cell Biol}\ }\textbf {\bibinfo {volume}
  {20}},\ \bibinfo {pages} {1118} (\bibinfo {year} {2018})}\BibitemShut
  {NoStop}%
\bibitem [{\citenamefont {Guo}\ \emph {et~al.}(2014)\citenamefont {Guo},
  \citenamefont {Ehrlicher}, \citenamefont {Jensen}, \citenamefont {Renz},
  \citenamefont {Moore}, \citenamefont {Goldman}, \citenamefont
  {Lippincott-Schwartz}, \citenamefont {Mackintosh},\ and\ \citenamefont
  {Weitz}}]{guo2014probing}%
  \BibitemOpen
  \bibfield  {author} {\bibinfo {author} {\bibfnamefont {M.}~\bibnamefont
  {Guo}}, \bibinfo {author} {\bibfnamefont {A.~J.}\ \bibnamefont {Ehrlicher}},
  \bibinfo {author} {\bibfnamefont {M.~H.}\ \bibnamefont {Jensen}}, \bibinfo
  {author} {\bibfnamefont {M.}~\bibnamefont {Renz}}, \bibinfo {author}
  {\bibfnamefont {J.~R.}\ \bibnamefont {Moore}}, \bibinfo {author}
  {\bibfnamefont {R.~D.}\ \bibnamefont {Goldman}}, \bibinfo {author}
  {\bibfnamefont {J.}~\bibnamefont {Lippincott-Schwartz}}, \bibinfo {author}
  {\bibfnamefont {F.~C.}\ \bibnamefont {Mackintosh}}, \ and\ \bibinfo {author}
  {\bibfnamefont {D.~A.}\ \bibnamefont {Weitz}},\ }\href@noop {} {\bibfield
  {journal} {\bibinfo  {journal} {Cell}\ }\textbf {\bibinfo {volume} {158}},\
  \bibinfo {pages} {822} (\bibinfo {year} {2014})}\BibitemShut {NoStop}%
\bibitem [{\citenamefont {Gal}\ \emph {et~al.}(2013)\citenamefont {Gal},
  \citenamefont {Lechtman-Goldstein},\ and\ \citenamefont
  {Weihs}}]{gal2013particle}%
  \BibitemOpen
  \bibfield  {author} {\bibinfo {author} {\bibfnamefont {N.}~\bibnamefont
  {Gal}}, \bibinfo {author} {\bibfnamefont {D.}~\bibnamefont
  {Lechtman-Goldstein}}, \ and\ \bibinfo {author} {\bibfnamefont
  {D.}~\bibnamefont {Weihs}},\ }\href@noop {} {\bibfield  {journal} {\bibinfo
  {journal} {Rheol Acta}\ }\textbf {\bibinfo {volume} {52}},\ \bibinfo {pages}
  {425} (\bibinfo {year} {2013})}\BibitemShut {NoStop}%
\bibitem [{\citenamefont {Koslover}\ \emph {et~al.}(2016)\citenamefont
  {Koslover}, \citenamefont {Chan},\ and\ \citenamefont
  {Theriot}}]{koslover2016disentangling}%
  \BibitemOpen
  \bibfield  {author} {\bibinfo {author} {\bibfnamefont {E.~F.}\ \bibnamefont
  {Koslover}}, \bibinfo {author} {\bibfnamefont {C.~K.}\ \bibnamefont {Chan}},
  \ and\ \bibinfo {author} {\bibfnamefont {J.~A.}\ \bibnamefont {Theriot}},\
  }\href@noop {} {\bibfield  {journal} {\bibinfo  {journal} {Biophys J}\
  }\textbf {\bibinfo {volume} {110}},\ \bibinfo {pages} {700} (\bibinfo {year}
  {2016})}\BibitemShut {NoStop}%
\bibitem [{\citenamefont {Lin}\ \emph {et~al.}(2016)\citenamefont {Lin},
  \citenamefont {Schuster}, \citenamefont {Guimaraes}, \citenamefont {Ashwin},
  \citenamefont {Schrader}, \citenamefont {Metz}, \citenamefont {Hacker},
  \citenamefont {Gurr},\ and\ \citenamefont {Steinberg}}]{lin2016active}%
  \BibitemOpen
  \bibfield  {author} {\bibinfo {author} {\bibfnamefont {C.}~\bibnamefont
  {Lin}}, \bibinfo {author} {\bibfnamefont {M.}~\bibnamefont {Schuster}},
  \bibinfo {author} {\bibfnamefont {S.~C.}\ \bibnamefont {Guimaraes}}, \bibinfo
  {author} {\bibfnamefont {P.}~\bibnamefont {Ashwin}}, \bibinfo {author}
  {\bibfnamefont {M.}~\bibnamefont {Schrader}}, \bibinfo {author}
  {\bibfnamefont {J.}~\bibnamefont {Metz}}, \bibinfo {author} {\bibfnamefont
  {C.}~\bibnamefont {Hacker}}, \bibinfo {author} {\bibfnamefont {S.~J.}\
  \bibnamefont {Gurr}}, \ and\ \bibinfo {author} {\bibfnamefont
  {G.}~\bibnamefont {Steinberg}},\ }\href@noop {} {\bibfield  {journal}
  {\bibinfo  {journal} {Nature communications}\ }\textbf {\bibinfo {volume}
  {7}},\ \bibinfo {pages} {11814} (\bibinfo {year} {2016})}\BibitemShut
  {NoStop}%
\bibitem [{\citenamefont {Chen}\ \emph {et~al.}(2015)\citenamefont {Chen},
  \citenamefont {Wang},\ and\ \citenamefont {Granick}}]{chen2015memoryless}%
  \BibitemOpen
  \bibfield  {author} {\bibinfo {author} {\bibfnamefont {K.}~\bibnamefont
  {Chen}}, \bibinfo {author} {\bibfnamefont {B.}~\bibnamefont {Wang}}, \ and\
  \bibinfo {author} {\bibfnamefont {S.}~\bibnamefont {Granick}},\ }\href@noop
  {} {\bibfield  {journal} {\bibinfo  {journal} {Nat Mater}\ }\textbf {\bibinfo
  {volume} {14}},\ \bibinfo {pages} {589} (\bibinfo {year} {2015})}\BibitemShut
  {NoStop}%
\bibitem [{\citenamefont {B{\'a}lint}\ \emph {et~al.}(2013)\citenamefont
  {B{\'a}lint}, \citenamefont {Vilanova}, \citenamefont {{\'A}lvarez},\ and\
  \citenamefont {Lakadamyali}}]{balint2013correlative}%
  \BibitemOpen
  \bibfield  {author} {\bibinfo {author} {\bibfnamefont {{\v{S}}.}~\bibnamefont
  {B{\'a}lint}}, \bibinfo {author} {\bibfnamefont {I.~V.}\ \bibnamefont
  {Vilanova}}, \bibinfo {author} {\bibfnamefont {{\'A}.~S.}\ \bibnamefont
  {{\'A}lvarez}}, \ and\ \bibinfo {author} {\bibfnamefont {M.}~\bibnamefont
  {Lakadamyali}},\ }\href@noop {} {\bibfield  {journal} {\bibinfo  {journal} {P
  Natl Acad Sci}\ }\textbf {\bibinfo {volume} {110}},\ \bibinfo {pages} {3375}
  (\bibinfo {year} {2013})}\BibitemShut {NoStop}%
\bibitem [{\citenamefont {Weihs}\ \emph {et~al.}(2006)\citenamefont {Weihs},
  \citenamefont {Mason},\ and\ \citenamefont {Teitell}}]{weihs2006bio}%
  \BibitemOpen
  \bibfield  {author} {\bibinfo {author} {\bibfnamefont {D.}~\bibnamefont
  {Weihs}}, \bibinfo {author} {\bibfnamefont {T.~G.}\ \bibnamefont {Mason}}, \
  and\ \bibinfo {author} {\bibfnamefont {M.~A.}\ \bibnamefont {Teitell}},\
  }\href@noop {} {\bibfield  {journal} {\bibinfo  {journal} {Biophys J}\
  }\textbf {\bibinfo {volume} {91}},\ \bibinfo {pages} {4296} (\bibinfo {year}
  {2006})}\BibitemShut {NoStop}%
\bibitem [{\citenamefont {Weber}\ \emph
  {et~al.}(2012{\natexlab{a}})\citenamefont {Weber}, \citenamefont
  {Spakowitz},\ and\ \citenamefont {Theriot}}]{weber2012nonthermal}%
  \BibitemOpen
  \bibfield  {author} {\bibinfo {author} {\bibfnamefont {S.~C.}\ \bibnamefont
  {Weber}}, \bibinfo {author} {\bibfnamefont {A.~J.}\ \bibnamefont
  {Spakowitz}}, \ and\ \bibinfo {author} {\bibfnamefont {J.~A.}\ \bibnamefont
  {Theriot}},\ }\href@noop {} {\bibfield  {journal} {\bibinfo  {journal} {P
  Natl Acad Sci}\ }\textbf {\bibinfo {volume} {109}},\ \bibinfo {pages} {7338}
  (\bibinfo {year} {2012}{\natexlab{a}})}\BibitemShut {NoStop}%
\bibitem [{\citenamefont {Fodor}\ \emph {et~al.}(2015)\citenamefont {Fodor},
  \citenamefont {Guo}, \citenamefont {Gov}, \citenamefont {Visco},
  \citenamefont {Weitz},\ and\ \citenamefont {van
  Wijland}}]{fodor2015activity}%
  \BibitemOpen
  \bibfield  {author} {\bibinfo {author} {\bibfnamefont {{\'E}.}~\bibnamefont
  {Fodor}}, \bibinfo {author} {\bibfnamefont {M.}~\bibnamefont {Guo}}, \bibinfo
  {author} {\bibfnamefont {N.}~\bibnamefont {Gov}}, \bibinfo {author}
  {\bibfnamefont {P.}~\bibnamefont {Visco}}, \bibinfo {author} {\bibfnamefont
  {D.}~\bibnamefont {Weitz}}, \ and\ \bibinfo {author} {\bibfnamefont
  {F.}~\bibnamefont {van Wijland}},\ }\href@noop {} {\bibfield  {journal}
  {\bibinfo  {journal} {EPL (Europhysics Letters)}\ }\textbf {\bibinfo {volume}
  {110}},\ \bibinfo {pages} {48005} (\bibinfo {year} {2015})}\BibitemShut
  {NoStop}%
\bibitem [{\citenamefont {Kollmannsberger}\ and\ \citenamefont
  {Fabry}(2011)}]{kollmannsberger2011linear}%
  \BibitemOpen
  \bibfield  {author} {\bibinfo {author} {\bibfnamefont {P.}~\bibnamefont
  {Kollmannsberger}}\ and\ \bibinfo {author} {\bibfnamefont {B.}~\bibnamefont
  {Fabry}},\ }\href@noop {} {\bibfield  {journal} {\bibinfo  {journal} {Annu
  Rev Mater Res}\ }\textbf {\bibinfo {volume} {41}},\ \bibinfo {pages} {75}
  (\bibinfo {year} {2011})}\BibitemShut {NoStop}%
\bibitem [{\citenamefont {Wang}\ \emph {et~al.}(2006)\citenamefont {Wang},
  \citenamefont {Austin},\ and\ \citenamefont {Cox}}]{wang2006single}%
  \BibitemOpen
  \bibfield  {author} {\bibinfo {author} {\bibfnamefont {Y.}~\bibnamefont
  {Wang}}, \bibinfo {author} {\bibfnamefont {R.~H.}\ \bibnamefont {Austin}}, \
  and\ \bibinfo {author} {\bibfnamefont {E.~C.}\ \bibnamefont {Cox}},\
  }\href@noop {} {\bibfield  {journal} {\bibinfo  {journal} {Phys Rev Lett}\
  }\textbf {\bibinfo {volume} {97}},\ \bibinfo {pages} {048302} (\bibinfo
  {year} {2006})}\BibitemShut {NoStop}%
\bibitem [{\citenamefont {Li}\ \emph {et~al.}(2015)\citenamefont {Li},
  \citenamefont {Dou}, \citenamefont {Liu}, \citenamefont {Li}, \citenamefont
  {Xie}, \citenamefont {Wang},\ and\ \citenamefont {Wang}}]{li2015mapping}%
  \BibitemOpen
  \bibfield  {author} {\bibinfo {author} {\bibfnamefont {H.}~\bibnamefont
  {Li}}, \bibinfo {author} {\bibfnamefont {S.-X.}\ \bibnamefont {Dou}},
  \bibinfo {author} {\bibfnamefont {Y.-R.}\ \bibnamefont {Liu}}, \bibinfo
  {author} {\bibfnamefont {W.}~\bibnamefont {Li}}, \bibinfo {author}
  {\bibfnamefont {P.}~\bibnamefont {Xie}}, \bibinfo {author} {\bibfnamefont
  {W.-C.}\ \bibnamefont {Wang}}, \ and\ \bibinfo {author} {\bibfnamefont
  {P.-Y.}\ \bibnamefont {Wang}},\ }\href@noop {} {\bibfield  {journal}
  {\bibinfo  {journal} {J Am Chem Soc}\ }\textbf {\bibinfo {volume} {137}},\
  \bibinfo {pages} {436} (\bibinfo {year} {2015})}\BibitemShut {NoStop}%
\bibitem [{\citenamefont {Etoc}\ \emph {et~al.}(2018)\citenamefont {Etoc},
  \citenamefont {Balloul}, \citenamefont {Vicario}, \citenamefont {Normanno},
  \citenamefont {Li{\ss}e}, \citenamefont {Sittner}, \citenamefont {Piehler},
  \citenamefont {Dahan},\ and\ \citenamefont {Coppey}}]{etoc2018non}%
  \BibitemOpen
  \bibfield  {author} {\bibinfo {author} {\bibfnamefont {F.}~\bibnamefont
  {Etoc}}, \bibinfo {author} {\bibfnamefont {E.}~\bibnamefont {Balloul}},
  \bibinfo {author} {\bibfnamefont {C.}~\bibnamefont {Vicario}}, \bibinfo
  {author} {\bibfnamefont {D.}~\bibnamefont {Normanno}}, \bibinfo {author}
  {\bibfnamefont {D.}~\bibnamefont {Li{\ss}e}}, \bibinfo {author}
  {\bibfnamefont {A.}~\bibnamefont {Sittner}}, \bibinfo {author} {\bibfnamefont
  {J.}~\bibnamefont {Piehler}}, \bibinfo {author} {\bibfnamefont
  {M.}~\bibnamefont {Dahan}}, \ and\ \bibinfo {author} {\bibfnamefont
  {M.}~\bibnamefont {Coppey}},\ }\href@noop {} {\bibfield  {journal} {\bibinfo
  {journal} {Nat Mater}\ }\textbf {\bibinfo {volume} {17}},\ \bibinfo {pages}
  {740} (\bibinfo {year} {2018})}\BibitemShut {NoStop}%
\bibitem [{\citenamefont {Yamada}\ \emph {et~al.}(2000)\citenamefont {Yamada},
  \citenamefont {Wirtz},\ and\ \citenamefont {Kuo}}]{yamada2000mechanics}%
  \BibitemOpen
  \bibfield  {author} {\bibinfo {author} {\bibfnamefont {S.}~\bibnamefont
  {Yamada}}, \bibinfo {author} {\bibfnamefont {D.}~\bibnamefont {Wirtz}}, \
  and\ \bibinfo {author} {\bibfnamefont {S.~C.}\ \bibnamefont {Kuo}},\
  }\href@noop {} {\bibfield  {journal} {\bibinfo  {journal} {Biophys J}\
  }\textbf {\bibinfo {volume} {78}},\ \bibinfo {pages} {1736} (\bibinfo {year}
  {2000})}\BibitemShut {NoStop}%
\bibitem [{\citenamefont {Hoffman}\ \emph {et~al.}(2006)\citenamefont
  {Hoffman}, \citenamefont {Massiera}, \citenamefont {Van~Citters},\ and\
  \citenamefont {Crocker}}]{hoffman2006consensus}%
  \BibitemOpen
  \bibfield  {author} {\bibinfo {author} {\bibfnamefont {B.~D.}\ \bibnamefont
  {Hoffman}}, \bibinfo {author} {\bibfnamefont {G.}~\bibnamefont {Massiera}},
  \bibinfo {author} {\bibfnamefont {K.~M.}\ \bibnamefont {Van~Citters}}, \ and\
  \bibinfo {author} {\bibfnamefont {J.~C.}\ \bibnamefont {Crocker}},\
  }\href@noop {} {\bibfield  {journal} {\bibinfo  {journal} {P Natl Acad Sci}\
  }\textbf {\bibinfo {volume} {103}},\ \bibinfo {pages} {10259} (\bibinfo
  {year} {2006})}\BibitemShut {NoStop}%
\bibitem [{\citenamefont {Weber}\ \emph
  {et~al.}(2012{\natexlab{b}})\citenamefont {Weber}, \citenamefont {Thompson},
  \citenamefont {Moerner}, \citenamefont {Spakowitz},\ and\ \citenamefont
  {Theriot}}]{weber2012analytical}%
  \BibitemOpen
  \bibfield  {author} {\bibinfo {author} {\bibfnamefont {S.~C.}\ \bibnamefont
  {Weber}}, \bibinfo {author} {\bibfnamefont {M.~A.}\ \bibnamefont {Thompson}},
  \bibinfo {author} {\bibfnamefont {W.~E.}\ \bibnamefont {Moerner}}, \bibinfo
  {author} {\bibfnamefont {A.~J.}\ \bibnamefont {Spakowitz}}, \ and\ \bibinfo
  {author} {\bibfnamefont {J.~A.}\ \bibnamefont {Theriot}},\ }\href@noop {}
  {\bibfield  {journal} {\bibinfo  {journal} {Biophys J}\ }\textbf {\bibinfo
  {volume} {102}},\ \bibinfo {pages} {2443} (\bibinfo {year}
  {2012}{\natexlab{b}})}\BibitemShut {NoStop}%
\bibitem [{\citenamefont {Sabri}\ \emph {et~al.}(2020)\citenamefont {Sabri},
  \citenamefont {Xu}, \citenamefont {Krapf},\ and\ \citenamefont
  {Weiss}}]{sabri2020elucidating}%
  \BibitemOpen
  \bibfield  {author} {\bibinfo {author} {\bibfnamefont {A.}~\bibnamefont
  {Sabri}}, \bibinfo {author} {\bibfnamefont {X.}~\bibnamefont {Xu}}, \bibinfo
  {author} {\bibfnamefont {D.}~\bibnamefont {Krapf}}, \ and\ \bibinfo {author}
  {\bibfnamefont {M.}~\bibnamefont {Weiss}},\ }\href@noop {} {\bibfield
  {journal} {\bibinfo  {journal} {Phys Rev Lett}\ }\textbf {\bibinfo {volume}
  {125}},\ \bibinfo {pages} {058101} (\bibinfo {year} {2020})}\BibitemShut
  {NoStop}%
\bibitem [{\citenamefont {Lucas}\ \emph {et~al.}(2014)\citenamefont {Lucas},
  \citenamefont {Zhang}, \citenamefont {Dudko},\ and\ \citenamefont
  {Murre}}]{lucas20143d}%
  \BibitemOpen
  \bibfield  {author} {\bibinfo {author} {\bibfnamefont {J.~S.}\ \bibnamefont
  {Lucas}}, \bibinfo {author} {\bibfnamefont {Y.}~\bibnamefont {Zhang}},
  \bibinfo {author} {\bibfnamefont {O.~K.}\ \bibnamefont {Dudko}}, \ and\
  \bibinfo {author} {\bibfnamefont {C.}~\bibnamefont {Murre}},\ }\href@noop {}
  {\bibfield  {journal} {\bibinfo  {journal} {Cell}\ }\textbf {\bibinfo
  {volume} {158}},\ \bibinfo {pages} {339} (\bibinfo {year}
  {2014})}\BibitemShut {NoStop}%
\bibitem [{\citenamefont {Wang}\ \emph {et~al.}(2012)\citenamefont {Wang},
  \citenamefont {Kuo}, \citenamefont {Bae},\ and\ \citenamefont
  {Granick}}]{wang2012brownian}%
  \BibitemOpen
  \bibfield  {author} {\bibinfo {author} {\bibfnamefont {B.}~\bibnamefont
  {Wang}}, \bibinfo {author} {\bibfnamefont {J.}~\bibnamefont {Kuo}}, \bibinfo
  {author} {\bibfnamefont {S.~C.}\ \bibnamefont {Bae}}, \ and\ \bibinfo
  {author} {\bibfnamefont {S.}~\bibnamefont {Granick}},\ }\href@noop {}
  {\bibfield  {journal} {\bibinfo  {journal} {Nat Mater}\ }\textbf {\bibinfo
  {volume} {11}},\ \bibinfo {pages} {481} (\bibinfo {year} {2012})}\BibitemShut
  {NoStop}%
\bibitem [{\citenamefont {Lampo}\ \emph {et~al.}(2017)\citenamefont {Lampo},
  \citenamefont {Stylianidou}, \citenamefont {Backlund}, \citenamefont
  {Wiggins},\ and\ \citenamefont {Spakowitz}}]{lampo2017cytoplasmic}%
  \BibitemOpen
  \bibfield  {author} {\bibinfo {author} {\bibfnamefont {T.~J.}\ \bibnamefont
  {Lampo}}, \bibinfo {author} {\bibfnamefont {S.}~\bibnamefont {Stylianidou}},
  \bibinfo {author} {\bibfnamefont {M.~P.}\ \bibnamefont {Backlund}}, \bibinfo
  {author} {\bibfnamefont {P.~A.}\ \bibnamefont {Wiggins}}, \ and\ \bibinfo
  {author} {\bibfnamefont {A.~J.}\ \bibnamefont {Spakowitz}},\ }\href@noop {}
  {\bibfield  {journal} {\bibinfo  {journal} {Biophys J}\ }\textbf {\bibinfo
  {volume} {112}},\ \bibinfo {pages} {532} (\bibinfo {year}
  {2017})}\BibitemShut {NoStop}%
\bibitem [{\citenamefont {Mogre}\ \emph {et~al.}(2020)\citenamefont {Mogre},
  \citenamefont {Brown},\ and\ \citenamefont {Koslover}}]{mogre2020getting}%
  \BibitemOpen
  \bibfield  {author} {\bibinfo {author} {\bibfnamefont {S.~S.}\ \bibnamefont
  {Mogre}}, \bibinfo {author} {\bibfnamefont {A.~I.}\ \bibnamefont {Brown}}, \
  and\ \bibinfo {author} {\bibfnamefont {E.~F.}\ \bibnamefont {Koslover}},\
  }\href@noop {} {\bibfield  {journal} {\bibinfo  {journal} {Phys Biol}\
  }\textbf {\bibinfo {volume} {17}},\ \bibinfo {pages} {061003} (\bibinfo
  {year} {2020})}\BibitemShut {NoStop}%
\bibitem [{\citenamefont {Duits}\ \emph {et~al.}(2009)\citenamefont {Duits},
  \citenamefont {Li}, \citenamefont {Vanapalli},\ and\ \citenamefont
  {Mugele}}]{duits2009mapping}%
  \BibitemOpen
  \bibfield  {author} {\bibinfo {author} {\bibfnamefont {M.~H.}\ \bibnamefont
  {Duits}}, \bibinfo {author} {\bibfnamefont {Y.}~\bibnamefont {Li}}, \bibinfo
  {author} {\bibfnamefont {S.~A.}\ \bibnamefont {Vanapalli}}, \ and\ \bibinfo
  {author} {\bibfnamefont {F.}~\bibnamefont {Mugele}},\ }\href@noop {}
  {\bibfield  {journal} {\bibinfo  {journal} {Phys Rev E}\ }\textbf {\bibinfo
  {volume} {79}},\ \bibinfo {pages} {051910} (\bibinfo {year}
  {2009})}\BibitemShut {NoStop}%
\bibitem [{\citenamefont {Chechkin}\ \emph {et~al.}(2017)\citenamefont
  {Chechkin}, \citenamefont {Seno}, \citenamefont {Metzler},\ and\
  \citenamefont {Sokolov}}]{chechkin2017brownian}%
  \BibitemOpen
  \bibfield  {author} {\bibinfo {author} {\bibfnamefont {A.~V.}\ \bibnamefont
  {Chechkin}}, \bibinfo {author} {\bibfnamefont {F.}~\bibnamefont {Seno}},
  \bibinfo {author} {\bibfnamefont {R.}~\bibnamefont {Metzler}}, \ and\
  \bibinfo {author} {\bibfnamefont {I.~M.}\ \bibnamefont {Sokolov}},\
  }\href@noop {} {\bibfield  {journal} {\bibinfo  {journal} {Physical Review
  X}\ }\textbf {\bibinfo {volume} {7}},\ \bibinfo {pages} {021002} (\bibinfo
  {year} {2017})}\BibitemShut {NoStop}%
\bibitem [{\citenamefont {Adler}\ \emph {et~al.}(2019)\citenamefont {Adler},
  \citenamefont {Sintorn}, \citenamefont {Strand},\ and\ \citenamefont
  {Parmryd}}]{adler2019conventional}%
  \BibitemOpen
  \bibfield  {author} {\bibinfo {author} {\bibfnamefont {J.}~\bibnamefont
  {Adler}}, \bibinfo {author} {\bibfnamefont {I.-M.}\ \bibnamefont {Sintorn}},
  \bibinfo {author} {\bibfnamefont {R.}~\bibnamefont {Strand}}, \ and\ \bibinfo
  {author} {\bibfnamefont {I.}~\bibnamefont {Parmryd}},\ }\href@noop {}
  {\bibfield  {journal} {\bibinfo  {journal} {Communications biology}\ }\textbf
  {\bibinfo {volume} {2}},\ \bibinfo {pages} {1} (\bibinfo {year}
  {2019})}\BibitemShut {NoStop}%
\bibitem [{\citenamefont {Saxton}(1995)}]{saxton1995single}%
  \BibitemOpen
  \bibfield  {author} {\bibinfo {author} {\bibfnamefont {M.~J.}\ \bibnamefont
  {Saxton}},\ }\href@noop {} {\bibfield  {journal} {\bibinfo  {journal}
  {Biophys J}\ }\textbf {\bibinfo {volume} {69}},\ \bibinfo {pages} {389}
  (\bibinfo {year} {1995})}\BibitemShut {NoStop}%
\bibitem [{\citenamefont {Viana}\ \emph {et~al.}(2020)\citenamefont {Viana},
  \citenamefont {Brown}, \citenamefont {Mueller}, \citenamefont {Goul},
  \citenamefont {Koslover},\ and\ \citenamefont
  {Rafelski}}]{viana2020mitochondrial}%
  \BibitemOpen
  \bibfield  {author} {\bibinfo {author} {\bibfnamefont {M.~P.}\ \bibnamefont
  {Viana}}, \bibinfo {author} {\bibfnamefont {A.~I.}\ \bibnamefont {Brown}},
  \bibinfo {author} {\bibfnamefont {I.~A.}\ \bibnamefont {Mueller}}, \bibinfo
  {author} {\bibfnamefont {C.}~\bibnamefont {Goul}}, \bibinfo {author}
  {\bibfnamefont {E.~F.}\ \bibnamefont {Koslover}}, \ and\ \bibinfo {author}
  {\bibfnamefont {S.~M.}\ \bibnamefont {Rafelski}},\ }\href {\doibase
  10.1016/j.cels.2020.02.002} {\bibfield  {journal} {\bibinfo  {journal} {Cell
  Syst}\ }\textbf {\bibinfo {volume} {10}},\ \bibinfo {pages} {287} (\bibinfo
  {year} {2020})}\BibitemShut {NoStop}%
\bibitem [{\citenamefont {Schwarz}\ and\ \citenamefont
  {Blower}(2016)}]{schwarz2016endoplasmic}%
  \BibitemOpen
  \bibfield  {author} {\bibinfo {author} {\bibfnamefont {D.~S.}\ \bibnamefont
  {Schwarz}}\ and\ \bibinfo {author} {\bibfnamefont {M.~D.}\ \bibnamefont
  {Blower}},\ }\href@noop {} {\bibfield  {journal} {\bibinfo  {journal} {Cell
  Mol Life Sci}\ }\textbf {\bibinfo {volume} {73}},\ \bibinfo {pages} {79}
  (\bibinfo {year} {2016})}\BibitemShut {NoStop}%
\bibitem [{\citenamefont {Perkins}\ and\ \citenamefont
  {Allan}(2021)}]{perkins2021intertwined}%
  \BibitemOpen
  \bibfield  {author} {\bibinfo {author} {\bibfnamefont {H.~T.}\ \bibnamefont
  {Perkins}}\ and\ \bibinfo {author} {\bibfnamefont {V.}~\bibnamefont
  {Allan}},\ }\href@noop {} {\bibfield  {journal} {\bibinfo  {journal} {Cells}\
  }\textbf {\bibinfo {volume} {10}},\ \bibinfo {pages} {2341} (\bibinfo {year}
  {2021})}\BibitemShut {NoStop}%
\bibitem [{\citenamefont {Westrate}\ \emph {et~al.}(2015)\citenamefont
  {Westrate}, \citenamefont {Lee}, \citenamefont {Prinz},\ and\ \citenamefont
  {Voeltz}}]{westrate2015form}%
  \BibitemOpen
  \bibfield  {author} {\bibinfo {author} {\bibfnamefont {L.}~\bibnamefont
  {Westrate}}, \bibinfo {author} {\bibfnamefont {J.}~\bibnamefont {Lee}},
  \bibinfo {author} {\bibfnamefont {W.}~\bibnamefont {Prinz}}, \ and\ \bibinfo
  {author} {\bibfnamefont {G.}~\bibnamefont {Voeltz}},\ }\href@noop {}
  {\bibfield  {journal} {\bibinfo  {journal} {Annu Rev Biochem}\ }\textbf
  {\bibinfo {volume} {84}},\ \bibinfo {pages} {791} (\bibinfo {year}
  {2015})}\BibitemShut {NoStop}%
\bibitem [{\citenamefont {Friedman}\ and\ \citenamefont
  {Voeltz}(2011)}]{friedman2011er}%
  \BibitemOpen
  \bibfield  {author} {\bibinfo {author} {\bibfnamefont {J.~R.}\ \bibnamefont
  {Friedman}}\ and\ \bibinfo {author} {\bibfnamefont {G.~K.}\ \bibnamefont
  {Voeltz}},\ }\href@noop {} {\bibfield  {journal} {\bibinfo  {journal} {Trends
  Cell Biol}\ }\textbf {\bibinfo {volume} {21}},\ \bibinfo {pages} {709}
  (\bibinfo {year} {2011})}\BibitemShut {NoStop}%
\bibitem [{\citenamefont {Chen}\ \emph {et~al.}(2013)\citenamefont {Chen},
  \citenamefont {Novick},\ and\ \citenamefont {Ferro-Novick}}]{chen2013er}%
  \BibitemOpen
  \bibfield  {author} {\bibinfo {author} {\bibfnamefont {S.}~\bibnamefont
  {Chen}}, \bibinfo {author} {\bibfnamefont {P.}~\bibnamefont {Novick}}, \ and\
  \bibinfo {author} {\bibfnamefont {S.}~\bibnamefont {Ferro-Novick}},\
  }\href@noop {} {\bibfield  {journal} {\bibinfo  {journal} {Curr Opin Cell
  Biol}\ }\textbf {\bibinfo {volume} {25}},\ \bibinfo {pages} {428} (\bibinfo
  {year} {2013})}\BibitemShut {NoStop}%
\bibitem [{\citenamefont {Griffing}(2010)}]{griffing2010networking}%
  \BibitemOpen
  \bibfield  {author} {\bibinfo {author} {\bibfnamefont {L.~R.}\ \bibnamefont
  {Griffing}},\ }\href@noop {} {\bibfield  {journal} {\bibinfo  {journal}
  {Biochem Soc T}\ }\textbf {\bibinfo {volume} {38}},\ \bibinfo {pages} {747}
  (\bibinfo {year} {2010})}\BibitemShut {NoStop}%
\bibitem [{\citenamefont {Barlowe}\ and\ \citenamefont
  {Helenius}(2016)}]{barlowe2016cargo}%
  \BibitemOpen
  \bibfield  {author} {\bibinfo {author} {\bibfnamefont {C.}~\bibnamefont
  {Barlowe}}\ and\ \bibinfo {author} {\bibfnamefont {A.}~\bibnamefont
  {Helenius}},\ }\href@noop {} {\bibfield  {journal} {\bibinfo  {journal} {Annu
  Rev Cell Dev Bi}\ }\textbf {\bibinfo {volume} {32}},\ \bibinfo {pages} {197}
  (\bibinfo {year} {2016})}\BibitemShut {NoStop}%
\bibitem [{\citenamefont {Brown}\ \emph {et~al.}(2020)\citenamefont {Brown},
  \citenamefont {Westrate},\ and\ \citenamefont {Koslover}}]{brown2020impact}%
  \BibitemOpen
  \bibfield  {author} {\bibinfo {author} {\bibfnamefont {A.~I.}\ \bibnamefont
  {Brown}}, \bibinfo {author} {\bibfnamefont {L.~M.}\ \bibnamefont {Westrate}},
  \ and\ \bibinfo {author} {\bibfnamefont {E.~F.}\ \bibnamefont {Koslover}},\
  }\href@noop {} {\bibfield  {journal} {\bibinfo  {journal} {Sci Rep}\ }\textbf
  {\bibinfo {volume} {10}},\ \bibinfo {pages} {1} (\bibinfo {year}
  {2020})}\BibitemShut {NoStop}%
\bibitem [{\citenamefont {Scott}\ \emph {et~al.}(2021)\citenamefont {Scott},
  \citenamefont {Brown}, \citenamefont {Mogre}, \citenamefont {Westrate},\ and\
  \citenamefont {Koslover}}]{scott2021diffusive}%
  \BibitemOpen
  \bibfield  {author} {\bibinfo {author} {\bibfnamefont {Z.~C.}\ \bibnamefont
  {Scott}}, \bibinfo {author} {\bibfnamefont {A.~I.}\ \bibnamefont {Brown}},
  \bibinfo {author} {\bibfnamefont {S.~S.}\ \bibnamefont {Mogre}}, \bibinfo
  {author} {\bibfnamefont {L.~M.}\ \bibnamefont {Westrate}}, \ and\ \bibinfo
  {author} {\bibfnamefont {E.~F.}\ \bibnamefont {Koslover}},\ }\href@noop {}
  {\bibfield  {journal} {\bibinfo  {journal} {Eur Phys J E}\ }\textbf {\bibinfo
  {volume} {44}},\ \bibinfo {pages} {1} (\bibinfo {year} {2021})}\BibitemShut
  {NoStop}%
\bibitem [{\citenamefont {Stadler}\ \emph {et~al.}(2018)\citenamefont
  {Stadler}, \citenamefont {Speckner},\ and\ \citenamefont
  {Weiss}}]{stadler2018diffusion}%
  \BibitemOpen
  \bibfield  {author} {\bibinfo {author} {\bibfnamefont {L.}~\bibnamefont
  {Stadler}}, \bibinfo {author} {\bibfnamefont {K.}~\bibnamefont {Speckner}}, \
  and\ \bibinfo {author} {\bibfnamefont {M.}~\bibnamefont {Weiss}},\
  }\href@noop {} {\bibfield  {journal} {\bibinfo  {journal} {Biophys J}\
  }\textbf {\bibinfo {volume} {115}},\ \bibinfo {pages} {1552} (\bibinfo {year}
  {2018})}\BibitemShut {NoStop}%
\bibitem [{\citenamefont {Dora}\ and\ \citenamefont
  {Holcman}(2020)}]{dora2020active}%
  \BibitemOpen
  \bibfield  {author} {\bibinfo {author} {\bibfnamefont {M.}~\bibnamefont
  {Dora}}\ and\ \bibinfo {author} {\bibfnamefont {D.}~\bibnamefont {Holcman}},\
  }\href@noop {} {\bibfield  {journal} {\bibinfo  {journal} {Proceedings of the
  Royal Society B}\ }\textbf {\bibinfo {volume} {287}},\ \bibinfo {pages}
  {20200493} (\bibinfo {year} {2020})}\BibitemShut {NoStop}%
\bibitem [{\citenamefont {Dayel}\ \emph {et~al.}(1999)\citenamefont {Dayel},
  \citenamefont {Hom},\ and\ \citenamefont {Verkman}}]{dayel1999diffusion}%
  \BibitemOpen
  \bibfield  {author} {\bibinfo {author} {\bibfnamefont {M.~J.}\ \bibnamefont
  {Dayel}}, \bibinfo {author} {\bibfnamefont {E.~F.}\ \bibnamefont {Hom}}, \
  and\ \bibinfo {author} {\bibfnamefont {A.~S.}\ \bibnamefont {Verkman}},\
  }\href@noop {} {\bibfield  {journal} {\bibinfo  {journal} {Biophys J}\
  }\textbf {\bibinfo {volume} {76}},\ \bibinfo {pages} {2843} (\bibinfo {year}
  {1999})}\BibitemShut {NoStop}%
\bibitem [{\citenamefont {Siggia}\ \emph {et~al.}(2000)\citenamefont {Siggia},
  \citenamefont {Lippincott-Schwartz},\ and\ \citenamefont
  {Bekiranov}}]{siggia2000diffusion}%
  \BibitemOpen
  \bibfield  {author} {\bibinfo {author} {\bibfnamefont {E.~D.}\ \bibnamefont
  {Siggia}}, \bibinfo {author} {\bibfnamefont {J.}~\bibnamefont
  {Lippincott-Schwartz}}, \ and\ \bibinfo {author} {\bibfnamefont
  {S.}~\bibnamefont {Bekiranov}},\ }\href@noop {} {\bibfield  {journal}
  {\bibinfo  {journal} {Biophysical journal}\ }\textbf {\bibinfo {volume}
  {79}},\ \bibinfo {pages} {1761} (\bibinfo {year} {2000})}\BibitemShut
  {NoStop}%
\bibitem [{\citenamefont {Nehls}\ \emph {et~al.}(2000)\citenamefont {Nehls},
  \citenamefont {Snapp}, \citenamefont {Cole}, \citenamefont {Zaal},
  \citenamefont {Kenworthy}, \citenamefont {Roberts}, \citenamefont
  {Ellenberg}, \citenamefont {Presley}, \citenamefont {Siggia},\ and\
  \citenamefont {Lippincott-Schwartz}}]{nehls2000dynamics}%
  \BibitemOpen
  \bibfield  {author} {\bibinfo {author} {\bibfnamefont {S.}~\bibnamefont
  {Nehls}}, \bibinfo {author} {\bibfnamefont {E.~L.}\ \bibnamefont {Snapp}},
  \bibinfo {author} {\bibfnamefont {N.~B.}\ \bibnamefont {Cole}}, \bibinfo
  {author} {\bibfnamefont {K.~J.}\ \bibnamefont {Zaal}}, \bibinfo {author}
  {\bibfnamefont {A.~K.}\ \bibnamefont {Kenworthy}}, \bibinfo {author}
  {\bibfnamefont {T.~H.}\ \bibnamefont {Roberts}}, \bibinfo {author}
  {\bibfnamefont {J.}~\bibnamefont {Ellenberg}}, \bibinfo {author}
  {\bibfnamefont {J.~F.}\ \bibnamefont {Presley}}, \bibinfo {author}
  {\bibfnamefont {E.}~\bibnamefont {Siggia}}, \ and\ \bibinfo {author}
  {\bibfnamefont {J.}~\bibnamefont {Lippincott-Schwartz}},\ }\href@noop {}
  {\bibfield  {journal} {\bibinfo  {journal} {Nature cell biology}\ }\textbf
  {\bibinfo {volume} {2}},\ \bibinfo {pages} {288} (\bibinfo {year}
  {2000})}\BibitemShut {NoStop}%
\bibitem [{\citenamefont {Konno}\ \emph {et~al.}(2021)\citenamefont {Konno},
  \citenamefont {Parutto}, \citenamefont {Bailey}, \citenamefont {Dav{\`\i}},
  \citenamefont {Crapart}, \citenamefont {Awadelkareem}, \citenamefont
  {Hockings}, \citenamefont {Brown}, \citenamefont {Xiang}, \citenamefont
  {Agrawal} \emph {et~al.}}]{konno2021endoplasmic}%
  \BibitemOpen
  \bibfield  {author} {\bibinfo {author} {\bibfnamefont {T.}~\bibnamefont
  {Konno}}, \bibinfo {author} {\bibfnamefont {P.}~\bibnamefont {Parutto}},
  \bibinfo {author} {\bibfnamefont {D.~M.}\ \bibnamefont {Bailey}}, \bibinfo
  {author} {\bibfnamefont {V.}~\bibnamefont {Dav{\`\i}}}, \bibinfo {author}
  {\bibfnamefont {C.}~\bibnamefont {Crapart}}, \bibinfo {author} {\bibfnamefont
  {M.~A.}\ \bibnamefont {Awadelkareem}}, \bibinfo {author} {\bibfnamefont
  {C.}~\bibnamefont {Hockings}}, \bibinfo {author} {\bibfnamefont
  {A.}~\bibnamefont {Brown}}, \bibinfo {author} {\bibfnamefont {K.~M.}\
  \bibnamefont {Xiang}}, \bibinfo {author} {\bibfnamefont {A.}~\bibnamefont
  {Agrawal}},  \emph {et~al.},\ }\href@noop {} {\bibfield  {journal} {\bibinfo
  {journal} {bioRxiv}\ } (\bibinfo {year} {2021})}\BibitemShut {NoStop}%
\bibitem [{\citenamefont {Santamaria}\ \emph {et~al.}(2006)\citenamefont
  {Santamaria}, \citenamefont {Wils}, \citenamefont {De~Schutter},\ and\
  \citenamefont {Augustine}}]{santamaria2006anomalous}%
  \BibitemOpen
  \bibfield  {author} {\bibinfo {author} {\bibfnamefont {F.}~\bibnamefont
  {Santamaria}}, \bibinfo {author} {\bibfnamefont {S.}~\bibnamefont {Wils}},
  \bibinfo {author} {\bibfnamefont {E.}~\bibnamefont {De~Schutter}}, \ and\
  \bibinfo {author} {\bibfnamefont {G.~J.}\ \bibnamefont {Augustine}},\
  }\href@noop {} {\bibfield  {journal} {\bibinfo  {journal} {Neuron}\ }\textbf
  {\bibinfo {volume} {52}},\ \bibinfo {pages} {635} (\bibinfo {year}
  {2006})}\BibitemShut {NoStop}%
\bibitem [{\citenamefont {Sartori}\ \emph {et~al.}(2020)\citenamefont
  {Sartori}, \citenamefont {Hafner}, \citenamefont {Karimi}, \citenamefont
  {Nold}, \citenamefont {Fonkeu}, \citenamefont {Schuman},\ and\ \citenamefont
  {Tchumatchenko}}]{sartori2020statistical}%
  \BibitemOpen
  \bibfield  {author} {\bibinfo {author} {\bibfnamefont {F.}~\bibnamefont
  {Sartori}}, \bibinfo {author} {\bibfnamefont {A.-S.}\ \bibnamefont {Hafner}},
  \bibinfo {author} {\bibfnamefont {A.}~\bibnamefont {Karimi}}, \bibinfo
  {author} {\bibfnamefont {A.}~\bibnamefont {Nold}}, \bibinfo {author}
  {\bibfnamefont {Y.}~\bibnamefont {Fonkeu}}, \bibinfo {author} {\bibfnamefont
  {E.~M.}\ \bibnamefont {Schuman}}, \ and\ \bibinfo {author} {\bibfnamefont
  {T.}~\bibnamefont {Tchumatchenko}},\ }\href@noop {} {\bibfield  {journal}
  {\bibinfo  {journal} {Cell Rep}\ }\textbf {\bibinfo {volume} {33}},\ \bibinfo
  {pages} {108391} (\bibinfo {year} {2020})}\BibitemShut {NoStop}%
\bibitem [{\citenamefont {Deng}\ and\ \citenamefont
  {Barkai}(2009)}]{deng2009ergodic}%
  \BibitemOpen
  \bibfield  {author} {\bibinfo {author} {\bibfnamefont {W.}~\bibnamefont
  {Deng}}\ and\ \bibinfo {author} {\bibfnamefont {E.}~\bibnamefont {Barkai}},\
  }\href@noop {} {\bibfield  {journal} {\bibinfo  {journal} {Phys Rev E}\
  }\textbf {\bibinfo {volume} {79}},\ \bibinfo {pages} {011112} (\bibinfo
  {year} {2009})}\BibitemShut {NoStop}%
\bibitem [{\citenamefont {Jeon}\ and\ \citenamefont
  {Metzler}(2010)}]{jeon2010fractional}%
  \BibitemOpen
  \bibfield  {author} {\bibinfo {author} {\bibfnamefont {J.-H.}\ \bibnamefont
  {Jeon}}\ and\ \bibinfo {author} {\bibfnamefont {R.}~\bibnamefont {Metzler}},\
  }\href@noop {} {\bibfield  {journal} {\bibinfo  {journal} {Phys Rev E}\
  }\textbf {\bibinfo {volume} {81}},\ \bibinfo {pages} {021103} (\bibinfo
  {year} {2010})}\BibitemShut {NoStop}%
\bibitem [{\citenamefont {Stauffer}\ and\ \citenamefont
  {Aharony}(2018)}]{stauffer2018introduction}%
  \BibitemOpen
  \bibfield  {author} {\bibinfo {author} {\bibfnamefont {D.}~\bibnamefont
  {Stauffer}}\ and\ \bibinfo {author} {\bibfnamefont {A.}~\bibnamefont
  {Aharony}},\ }\href@noop {} {\emph {\bibinfo {title} {Introduction to
  percolation theory}}}\ (\bibinfo  {publisher} {CRC press},\ \bibinfo {year}
  {2018})\BibitemShut {NoStop}%
\bibitem [{\citenamefont {Ross}(2014)}]{ross2014introduction}%
  \BibitemOpen
  \bibfield  {author} {\bibinfo {author} {\bibfnamefont {S.~M.}\ \bibnamefont
  {Ross}},\ }\href@noop {} {\emph {\bibinfo {title} {Introduction to
  probability models}}}\ (\bibinfo  {publisher} {Academic press},\ \bibinfo
  {year} {2014})\BibitemShut {NoStop}%
\bibitem [{\citenamefont {Metzler}\ and\ \citenamefont
  {Klafter}(2000)}]{metzler2000random}%
  \BibitemOpen
  \bibfield  {author} {\bibinfo {author} {\bibfnamefont {R.}~\bibnamefont
  {Metzler}}\ and\ \bibinfo {author} {\bibfnamefont {J.}~\bibnamefont
  {Klafter}},\ }\href@noop {} {\bibfield  {journal} {\bibinfo  {journal}
  {Physics reports}\ }\textbf {\bibinfo {volume} {339}},\ \bibinfo {pages} {1}
  (\bibinfo {year} {2000})}\BibitemShut {NoStop}%
\bibitem [{\citenamefont {Weber}\ \emph
  {et~al.}(2010{\natexlab{a}})\citenamefont {Weber}, \citenamefont
  {Spakowitz},\ and\ \citenamefont {Theriot}}]{weber2010bacterial}%
  \BibitemOpen
  \bibfield  {author} {\bibinfo {author} {\bibfnamefont {S.~C.}\ \bibnamefont
  {Weber}}, \bibinfo {author} {\bibfnamefont {A.~J.}\ \bibnamefont
  {Spakowitz}}, \ and\ \bibinfo {author} {\bibfnamefont {J.~A.}\ \bibnamefont
  {Theriot}},\ }\href@noop {} {\bibfield  {journal} {\bibinfo  {journal} {Phys
  Rev Lett}\ }\textbf {\bibinfo {volume} {104}},\ \bibinfo {pages} {238102}
  (\bibinfo {year} {2010}{\natexlab{a}})}\BibitemShut {NoStop}%
\bibitem [{\citenamefont {Bressloff}\ and\ \citenamefont
  {Newby}(2013)}]{bressloff2013stochastic}%
  \BibitemOpen
  \bibfield  {author} {\bibinfo {author} {\bibfnamefont {P.~C.}\ \bibnamefont
  {Bressloff}}\ and\ \bibinfo {author} {\bibfnamefont {J.~M.}\ \bibnamefont
  {Newby}},\ }\href@noop {} {\bibfield  {journal} {\bibinfo  {journal} {Rev Mod
  Phys}\ }\textbf {\bibinfo {volume} {85}},\ \bibinfo {pages} {135} (\bibinfo
  {year} {2013})}\BibitemShut {NoStop}%
\bibitem [{\citenamefont {Kroese}\ and\ \citenamefont
  {Botev}(2013)}]{kroese2013spatial}%
  \BibitemOpen
  \bibfield  {author} {\bibinfo {author} {\bibfnamefont {D.}~\bibnamefont
  {Kroese}}\ and\ \bibinfo {author} {\bibfnamefont {Z.}~\bibnamefont {Botev}},\
  }\href@noop {} {\bibfield  {journal} {\bibinfo  {journal} {Analysis, modeling
  and simulation of complex structures}\ }\textbf {\bibinfo {volume} {2}}
  (\bibinfo {year} {2013})}\BibitemShut {NoStop}%
\bibitem [{\citenamefont {Weber}\ \emph
  {et~al.}(2010{\natexlab{b}})\citenamefont {Weber}, \citenamefont {Theriot},\
  and\ \citenamefont {Spakowitz}}]{weber2010subdiffusive}%
  \BibitemOpen
  \bibfield  {author} {\bibinfo {author} {\bibfnamefont {S.~C.}\ \bibnamefont
  {Weber}}, \bibinfo {author} {\bibfnamefont {J.~A.}\ \bibnamefont {Theriot}},
  \ and\ \bibinfo {author} {\bibfnamefont {A.~J.}\ \bibnamefont {Spakowitz}},\
  }\href@noop {} {\bibfield  {journal} {\bibinfo  {journal} {Phys Rev E}\
  }\textbf {\bibinfo {volume} {82}},\ \bibinfo {pages} {011913} (\bibinfo
  {year} {2010}{\natexlab{b}})}\BibitemShut {NoStop}%
\bibitem [{\citenamefont {Friedman}\ \emph {et~al.}(2010)\citenamefont
  {Friedman}, \citenamefont {Webster}, \citenamefont {Mastronarde},
  \citenamefont {Verhey},\ and\ \citenamefont {Voeltz}}]{friedman2010er}%
  \BibitemOpen
  \bibfield  {author} {\bibinfo {author} {\bibfnamefont {J.~R.}\ \bibnamefont
  {Friedman}}, \bibinfo {author} {\bibfnamefont {B.~M.}\ \bibnamefont
  {Webster}}, \bibinfo {author} {\bibfnamefont {D.~N.}\ \bibnamefont
  {Mastronarde}}, \bibinfo {author} {\bibfnamefont {K.~J.}\ \bibnamefont
  {Verhey}}, \ and\ \bibinfo {author} {\bibfnamefont {G.~K.}\ \bibnamefont
  {Voeltz}},\ }\href@noop {} {\bibfield  {journal} {\bibinfo  {journal} {J Cell
  Biol}\ }\textbf {\bibinfo {volume} {190}},\ \bibinfo {pages} {363} (\bibinfo
  {year} {2010})}\BibitemShut {NoStop}%
\bibitem [{\citenamefont {Lin}\ \emph {et~al.}(2017)\citenamefont {Lin},
  \citenamefont {White}, \citenamefont {Sparkes},\ and\ \citenamefont
  {Ashwin}}]{lin2017modeling}%
  \BibitemOpen
  \bibfield  {author} {\bibinfo {author} {\bibfnamefont {C.}~\bibnamefont
  {Lin}}, \bibinfo {author} {\bibfnamefont {R.~R.}\ \bibnamefont {White}},
  \bibinfo {author} {\bibfnamefont {I.}~\bibnamefont {Sparkes}}, \ and\
  \bibinfo {author} {\bibfnamefont {P.}~\bibnamefont {Ashwin}},\ }\href@noop {}
  {\bibfield  {journal} {\bibinfo  {journal} {Biophys J}\ }\textbf {\bibinfo
  {volume} {113}},\ \bibinfo {pages} {214} (\bibinfo {year}
  {2017})}\BibitemShut {NoStop}%
\bibitem [{\citenamefont {Lee}\ and\ \citenamefont
  {Chen}(1988)}]{lee1988dynamic}%
  \BibitemOpen
  \bibfield  {author} {\bibinfo {author} {\bibfnamefont {C.}~\bibnamefont
  {Lee}}\ and\ \bibinfo {author} {\bibfnamefont {L.~B.}\ \bibnamefont {Chen}},\
  }\href@noop {} {\bibfield  {journal} {\bibinfo  {journal} {Cell}\ }\textbf
  {\bibinfo {volume} {54}},\ \bibinfo {pages} {37} (\bibinfo {year}
  {1988})}\BibitemShut {NoStop}%
\bibitem [{\citenamefont {Lee}\ \emph {et~al.}(1989)\citenamefont {Lee},
  \citenamefont {Ferguson},\ and\ \citenamefont {Chen}}]{lee1989construction}%
  \BibitemOpen
  \bibfield  {author} {\bibinfo {author} {\bibfnamefont {C.}~\bibnamefont
  {Lee}}, \bibinfo {author} {\bibfnamefont {M.}~\bibnamefont {Ferguson}}, \
  and\ \bibinfo {author} {\bibfnamefont {L.~B.}\ \bibnamefont {Chen}},\
  }\href@noop {} {\bibfield  {journal} {\bibinfo  {journal} {The Journal of
  cell biology}\ }\textbf {\bibinfo {volume} {109}},\ \bibinfo {pages} {2045}
  (\bibinfo {year} {1989})}\BibitemShut {NoStop}%
\bibitem [{\citenamefont {Schroeder}\ \emph {et~al.}(2019)\citenamefont
  {Schroeder}, \citenamefont {Barentine}, \citenamefont {Merta}, \citenamefont
  {Schweighofer}, \citenamefont {Zhang}, \citenamefont {Baddeley},
  \citenamefont {Bewersdorf},\ and\ \citenamefont
  {Bahmanyar}}]{schroeder2019dynamic}%
  \BibitemOpen
  \bibfield  {author} {\bibinfo {author} {\bibfnamefont {L.~K.}\ \bibnamefont
  {Schroeder}}, \bibinfo {author} {\bibfnamefont {A.~E.}\ \bibnamefont
  {Barentine}}, \bibinfo {author} {\bibfnamefont {H.}~\bibnamefont {Merta}},
  \bibinfo {author} {\bibfnamefont {S.}~\bibnamefont {Schweighofer}}, \bibinfo
  {author} {\bibfnamefont {Y.}~\bibnamefont {Zhang}}, \bibinfo {author}
  {\bibfnamefont {D.}~\bibnamefont {Baddeley}}, \bibinfo {author}
  {\bibfnamefont {J.}~\bibnamefont {Bewersdorf}}, \ and\ \bibinfo {author}
  {\bibfnamefont {S.}~\bibnamefont {Bahmanyar}},\ }\href@noop {} {\bibfield
  {journal} {\bibinfo  {journal} {Journal of Cell Biology}\ }\textbf {\bibinfo
  {volume} {218}},\ \bibinfo {pages} {83} (\bibinfo {year} {2019})}\BibitemShut
  {NoStop}%
\bibitem [{\citenamefont {Barth{\'e}lemy}(2011)}]{barthelemy2011spatial}%
  \BibitemOpen
  \bibfield  {author} {\bibinfo {author} {\bibfnamefont {M.}~\bibnamefont
  {Barth{\'e}lemy}},\ }\href@noop {} {\bibfield  {journal} {\bibinfo  {journal}
  {Physics Reports}\ }\textbf {\bibinfo {volume} {499}},\ \bibinfo {pages} {1}
  (\bibinfo {year} {2011})}\BibitemShut {NoStop}%
\bibitem [{\citenamefont {Malchus}\ and\ \citenamefont
  {Weiss}(2010)}]{malchus2010anomalous}%
  \BibitemOpen
  \bibfield  {author} {\bibinfo {author} {\bibfnamefont {N.}~\bibnamefont
  {Malchus}}\ and\ \bibinfo {author} {\bibfnamefont {M.}~\bibnamefont
  {Weiss}},\ }\href@noop {} {\bibfield  {journal} {\bibinfo  {journal} {Biophys
  J}\ }\textbf {\bibinfo {volume} {99}},\ \bibinfo {pages} {1321} (\bibinfo
  {year} {2010})}\BibitemShut {NoStop}%
\bibitem [{\citenamefont {Wang}\ \emph {et~al.}(2016)\citenamefont {Wang},
  \citenamefont {Tukachinsky}, \citenamefont {Romano},\ and\ \citenamefont
  {Rapoport}}]{wang2016cooperation}%
  \BibitemOpen
  \bibfield  {author} {\bibinfo {author} {\bibfnamefont {S.}~\bibnamefont
  {Wang}}, \bibinfo {author} {\bibfnamefont {H.}~\bibnamefont {Tukachinsky}},
  \bibinfo {author} {\bibfnamefont {F.~B.}\ \bibnamefont {Romano}}, \ and\
  \bibinfo {author} {\bibfnamefont {T.~A.}\ \bibnamefont {Rapoport}},\
  }\href@noop {} {\bibfield  {journal} {\bibinfo  {journal} {{elife}}\ }\textbf
  {\bibinfo {volume} {5}},\ \bibinfo {pages} {e18605} (\bibinfo {year}
  {2016})}\BibitemShut {NoStop}%
\bibitem [{\citenamefont {MATLAB}(2018)}]{MATLAB:2018}%
  \BibitemOpen
  \bibfield  {author} {\bibinfo {author} {\bibnamefont {MATLAB}},\ }\href@noop
  {} {\emph {\bibinfo {title} {version 9.5 (R2018b)}}}\ (\bibinfo  {publisher}
  {The MathWorks Inc.},\ \bibinfo {address} {Natick, Massachusetts},\ \bibinfo
  {year} {2018})\BibitemShut {NoStop}%
\bibitem [{\citenamefont {Sengupta}\ \emph {et~al.}(2013)\citenamefont
  {Sengupta}, \citenamefont {Jovanovic-Talisman},\ and\ \citenamefont
  {Lippincott-Schwartz}}]{sengupta2013quantifying}%
  \BibitemOpen
  \bibfield  {author} {\bibinfo {author} {\bibfnamefont {P.}~\bibnamefont
  {Sengupta}}, \bibinfo {author} {\bibfnamefont {T.}~\bibnamefont
  {Jovanovic-Talisman}}, \ and\ \bibinfo {author} {\bibfnamefont
  {J.}~\bibnamefont {Lippincott-Schwartz}},\ }\href@noop {} {\bibfield
  {journal} {\bibinfo  {journal} {Nature protocols}\ }\textbf {\bibinfo
  {volume} {8}},\ \bibinfo {pages} {345} (\bibinfo {year} {2013})}\BibitemShut
  {NoStop}%
\bibitem [{\citenamefont {Nixon-Abell}\ \emph {et~al.}(2016)\citenamefont
  {Nixon-Abell}, \citenamefont {Obara}, \citenamefont {Weigel}, \citenamefont
  {Li}, \citenamefont {Legant}, \citenamefont {Xu}, \citenamefont {Pasolli},
  \citenamefont {Harvey}, \citenamefont {Hess}, \citenamefont {Betzig} \emph
  {et~al.}}]{nixon2016increased}%
  \BibitemOpen
  \bibfield  {author} {\bibinfo {author} {\bibfnamefont {J.}~\bibnamefont
  {Nixon-Abell}}, \bibinfo {author} {\bibfnamefont {C.~J.}\ \bibnamefont
  {Obara}}, \bibinfo {author} {\bibfnamefont {A.~V.}\ \bibnamefont {Weigel}},
  \bibinfo {author} {\bibfnamefont {D.}~\bibnamefont {Li}}, \bibinfo {author}
  {\bibfnamefont {W.~R.}\ \bibnamefont {Legant}}, \bibinfo {author}
  {\bibfnamefont {C.~S.}\ \bibnamefont {Xu}}, \bibinfo {author} {\bibfnamefont
  {H.~A.}\ \bibnamefont {Pasolli}}, \bibinfo {author} {\bibfnamefont
  {K.}~\bibnamefont {Harvey}}, \bibinfo {author} {\bibfnamefont {H.~F.}\
  \bibnamefont {Hess}}, \bibinfo {author} {\bibfnamefont {E.}~\bibnamefont
  {Betzig}},  \emph {et~al.},\ }\href@noop {} {\bibfield  {journal} {\bibinfo
  {journal} {Science}\ }\textbf {\bibinfo {volume} {354}} (\bibinfo {year}
  {2016})}\BibitemShut {NoStop}%
\bibitem [{\citenamefont {Grimm}\ \emph {et~al.}(2016)\citenamefont {Grimm},
  \citenamefont {English}, \citenamefont {Choi}, \citenamefont {Muthusamy},
  \citenamefont {Mehl}, \citenamefont {Dong}, \citenamefont {Brown},
  \citenamefont {Lippincott-Schwartz}, \citenamefont {Liu}, \citenamefont
  {Lionnet} \emph {et~al.}}]{grimm2016bright}%
  \BibitemOpen
  \bibfield  {author} {\bibinfo {author} {\bibfnamefont {J.~B.}\ \bibnamefont
  {Grimm}}, \bibinfo {author} {\bibfnamefont {B.~P.}\ \bibnamefont {English}},
  \bibinfo {author} {\bibfnamefont {H.}~\bibnamefont {Choi}}, \bibinfo {author}
  {\bibfnamefont {A.~K.}\ \bibnamefont {Muthusamy}}, \bibinfo {author}
  {\bibfnamefont {B.~P.}\ \bibnamefont {Mehl}}, \bibinfo {author}
  {\bibfnamefont {P.}~\bibnamefont {Dong}}, \bibinfo {author} {\bibfnamefont
  {T.~A.}\ \bibnamefont {Brown}}, \bibinfo {author} {\bibfnamefont
  {J.}~\bibnamefont {Lippincott-Schwartz}}, \bibinfo {author} {\bibfnamefont
  {Z.}~\bibnamefont {Liu}}, \bibinfo {author} {\bibfnamefont {T.}~\bibnamefont
  {Lionnet}},  \emph {et~al.},\ }\href@noop {} {\bibfield  {journal} {\bibinfo
  {journal} {Nature methods}\ }\textbf {\bibinfo {volume} {13}},\ \bibinfo
  {pages} {985} (\bibinfo {year} {2016})}\BibitemShut {NoStop}%
\bibitem [{\citenamefont {Berg}\ \emph {et~al.}(2019)\citenamefont {Berg},
  \citenamefont {Kutra}, \citenamefont {Kroeger}, \citenamefont {Straehle},
  \citenamefont {Kausler}, \citenamefont {Haubold}, \citenamefont {Schiegg},
  \citenamefont {Ales}, \citenamefont {Beier}, \citenamefont {Rudy},
  \citenamefont {Eren}, \citenamefont {Cervantes}, \citenamefont {Xu},
  \citenamefont {Beuttenmueller}, \citenamefont {Wolny}, \citenamefont {Zhang},
  \citenamefont {Koethe}, \citenamefont {Hamprecht},\ and\ \citenamefont
  {Kreshuk}}]{berg2019ilastik}%
  \BibitemOpen
  \bibfield  {author} {\bibinfo {author} {\bibfnamefont {S.}~\bibnamefont
  {Berg}}, \bibinfo {author} {\bibfnamefont {D.}~\bibnamefont {Kutra}},
  \bibinfo {author} {\bibfnamefont {T.}~\bibnamefont {Kroeger}}, \bibinfo
  {author} {\bibfnamefont {C.~N.}\ \bibnamefont {Straehle}}, \bibinfo {author}
  {\bibfnamefont {B.~X.}\ \bibnamefont {Kausler}}, \bibinfo {author}
  {\bibfnamefont {C.}~\bibnamefont {Haubold}}, \bibinfo {author} {\bibfnamefont
  {M.}~\bibnamefont {Schiegg}}, \bibinfo {author} {\bibfnamefont
  {J.}~\bibnamefont {Ales}}, \bibinfo {author} {\bibfnamefont {T.}~\bibnamefont
  {Beier}}, \bibinfo {author} {\bibfnamefont {M.}~\bibnamefont {Rudy}},
  \bibinfo {author} {\bibfnamefont {K.}~\bibnamefont {Eren}}, \bibinfo {author}
  {\bibfnamefont {J.~I.}\ \bibnamefont {Cervantes}}, \bibinfo {author}
  {\bibfnamefont {B.}~\bibnamefont {Xu}}, \bibinfo {author} {\bibfnamefont
  {F.}~\bibnamefont {Beuttenmueller}}, \bibinfo {author} {\bibfnamefont
  {A.}~\bibnamefont {Wolny}}, \bibinfo {author} {\bibfnamefont
  {C.}~\bibnamefont {Zhang}}, \bibinfo {author} {\bibfnamefont
  {U.}~\bibnamefont {Koethe}}, \bibinfo {author} {\bibfnamefont {F.~A.}\
  \bibnamefont {Hamprecht}}, \ and\ \bibinfo {author} {\bibfnamefont
  {A.}~\bibnamefont {Kreshuk}},\ }\href {\doibase 10.1038/s41592-019-0582-9}
  {\bibfield  {journal} {\bibinfo  {journal} {Nat Methods}\ } (\bibinfo {year}
  {2019}),\ 10.1038/s41592-019-0582-9}\BibitemShut {NoStop}%
\bibitem [{\citenamefont {Tinevez}\ \emph {et~al.}(2017)\citenamefont
  {Tinevez}, \citenamefont {Perry}, \citenamefont {Schindelin}, \citenamefont
  {Hoopes}, \citenamefont {Reynolds}, \citenamefont {Laplantine}, \citenamefont
  {Bednarek}, \citenamefont {Shorte},\ and\ \citenamefont
  {Eliceiri}}]{tinevez2017trackmate}%
  \BibitemOpen
  \bibfield  {author} {\bibinfo {author} {\bibfnamefont {J.-Y.}\ \bibnamefont
  {Tinevez}}, \bibinfo {author} {\bibfnamefont {N.}~\bibnamefont {Perry}},
  \bibinfo {author} {\bibfnamefont {J.}~\bibnamefont {Schindelin}}, \bibinfo
  {author} {\bibfnamefont {G.~M.}\ \bibnamefont {Hoopes}}, \bibinfo {author}
  {\bibfnamefont {G.~D.}\ \bibnamefont {Reynolds}}, \bibinfo {author}
  {\bibfnamefont {E.}~\bibnamefont {Laplantine}}, \bibinfo {author}
  {\bibfnamefont {S.~Y.}\ \bibnamefont {Bednarek}}, \bibinfo {author}
  {\bibfnamefont {S.~L.}\ \bibnamefont {Shorte}}, \ and\ \bibinfo {author}
  {\bibfnamefont {K.~W.}\ \bibnamefont {Eliceiri}},\ }\href@noop {} {\bibfield
  {journal} {\bibinfo  {journal} {Methods}\ }\textbf {\bibinfo {volume}
  {115}},\ \bibinfo {pages} {80} (\bibinfo {year} {2017})}\BibitemShut
  {NoStop}%
\bibitem [{\citenamefont {Jaqaman}\ \emph {et~al.}(2008)\citenamefont
  {Jaqaman}, \citenamefont {Loerke}, \citenamefont {Mettlen}, \citenamefont
  {Kuwata}, \citenamefont {Grinstein}, \citenamefont {Schmid},\ and\
  \citenamefont {Danuser}}]{jaqaman2008robust}%
  \BibitemOpen
  \bibfield  {author} {\bibinfo {author} {\bibfnamefont {K.}~\bibnamefont
  {Jaqaman}}, \bibinfo {author} {\bibfnamefont {D.}~\bibnamefont {Loerke}},
  \bibinfo {author} {\bibfnamefont {M.}~\bibnamefont {Mettlen}}, \bibinfo
  {author} {\bibfnamefont {H.}~\bibnamefont {Kuwata}}, \bibinfo {author}
  {\bibfnamefont {S.}~\bibnamefont {Grinstein}}, \bibinfo {author}
  {\bibfnamefont {S.~L.}\ \bibnamefont {Schmid}}, \ and\ \bibinfo {author}
  {\bibfnamefont {G.}~\bibnamefont {Danuser}},\ }\href@noop {} {\bibfield
  {journal} {\bibinfo  {journal} {Nature methods}\ }\textbf {\bibinfo {volume}
  {5}},\ \bibinfo {pages} {695} (\bibinfo {year} {2008})}\BibitemShut {NoStop}%
\end{thebibliography}%

\appendix
\section{Methods}
\label{app:methods}

All code used in this work was written in Matlab~\cite{MATLAB:2018}. A software package for carrying out simulations on networks and for analyzing particle trajectories will be provided at \url{https://github.com/lenafabr/unravelNetworkTraj} prior to publication.

\subsection{Algorithm for Brownian motion simulations on networks}
\label{sec:simalgorithm}
We develop an algorithm for discrete time-step simulations of Brownian particles on a network of one-dimensional edges connected by point-like junctions, as discussed in Sec.~\ref{sec:simdiff}.

At each step, for each individual particle, we consider the particle's position along its current edge. Without loss of generality, we can assume that the nearest boundary (node) of that edge is at position $0$ and the particle itself is at position $x_0$ along the edge. We sample the displacement of the particle using a normal distribution $\Delta z \sim \mathcal{N}(0,\sqrt{2D\Delta t})$. We then treat the particle as if it were on an infinite line and consider, of all possible paths that start at $x_0$ and end at $x_0+\Delta z$, what is the probability that the particle's path crossed $0$ during that timestep. This probability $p_\text{pass}$ is derived as follows.

We start with the Green's function for unconstrained diffusion:
\begin{equation}
	\begin{split}
	G(x|x_0,\Delta t) = \frac{1}{\sqrt{4\pi Dt}} e^{-(x-x_0)^2/{4Dt}},
	\end{split}
	\label{eq:G}
\end{equation}
which gives the spatial probability density of a particle ending its path at position $x$ after time $\Delta t$, given that it started at position $x_0$ at time $0$. The distribution of first passage times for a particle starting at $x_0$ to hit $0$ for the first time at time $t$ is given by
\begin{equation}
	\begin{split}
	J(t; x_0) = D \left. \frac{\partial}{\partial x} G(x|x_0; t)\right|_{x=0}.
	\end{split}
	\label{eq:J}
\end{equation}

Because diffusion is Markovian (memory-less), if the particle first hits $0$ at time $t$, we know the probability density of ending up at position $x$ is simply given by $G(x|0,\Delta t - t)$. 
From there, we can calculate the conditional probability that a particle hit $0$ at some point in the path, given that it ended at $x$, by a simple application of Bayes' rule:
\begin{equation}
	\begin{split}
	p_\text{pass} = \frac{\int_0^{\Delta t} J(t;x_0) G(x|0,\Delta t- t) dt}{G(x|x_0,\Delta t)}.	
	\end{split}
\end{equation}
Here, the numerator is the joint probability density of hitting $0$ at some point during the time-step and then ending the path at $x$ and the denominator is the overall probability density  of ending at $x$. Plugging in Eq.~\ref{eq:G} and \ref{eq:J} gives the formula for $p_\text{pass}$ stated in the main text (Eq.~\ref{eqn:ppass}).

Every path of an unconstrained particle can be mapped to a set of `folded' paths for the equivalent particle on the network. Each passage of the node at $0$ involves selecting which of the adjacent network edges will serve as the axis along which the network-bound particle continues to move. For a memory-less Brownian particle that hits the node, each of the edges (including the one from whence it came) are equally likely to be selected. We know the unconstrained particle ends its timestep at distance $|x_0+\Delta z|$ from $0$. Hence, the network-confined particle is placed at this distance from the node along the randomly selected edge, completing its time-step.

\subsection{Distribution of diffusing particles on a triskelion}
\label{app:triskelion}

We compute explicitly the solution to the diffusion equation on a triskelion structure (Fig.~\ref{fig:simMSD}b) with reflecting boundaries at the tips and continuity of the concentrations at the degree-3 node. The initial condition is assumed to be a delta function at the junction. The approach taken is analogous to the equivalent problem with absorbing boundaries, described in more detail in Ref.~\cite{scott2021diffusive}. Specifically, after a Laplace transform in time ($t\rightarrow s$), the concentration profile  $\hat{c}_k(x,s)$ on edge $k$ is given by
\begin{equation}
\begin{split}	
\hat{c}_k(x,s) = \frac{1}{D\alpha} \frac{\cosh \alpha x}{\cosh \alpha \ell_k} \left[\sum_{j=1}^3 \tanh \alpha \ell_j \right]^{-1}
\end{split}
\label{eq:triskelioncs}
\end{equation}
where $D$ is the particle diffusivity, $\alpha = \sqrt{s/D}$, and  $\ell_j$ is the length of the $j^\text{th}$ edge.

The Laplace transform is then inverted using a Bromwich integral evaluated with the Cauchy residue theorem, yielding the time-dependent concentration profile
\begin{equation}
\begin{split}	
c(x,t)	&= \sum_p r_p e^{-D u_p^2 t},
\end{split}
\end{equation}
where the poles $s_p = -D u_p^2$ can be found by taking the roots of the following equation,
\begin{equation}
\begin{split}
	\cos\ell_k u_p & \left[\sum_j \tan \ell_j u_p \right] = 0,
\end{split}
\end{equation}
and $r_p$ are the residues of Eq.~\ref{eq:triskelioncs} at those poles.

The resulting expressions are plotted in Fig.~\ref{fig:simMSD}b and compared to the distribution of $10^7$ simulated particles at time $t = 0.8\ell^2/D$ (after starting at the three-fold junction) where $\ell$ is the length of the shortest edge. 

\subsection{Parameters for simulations and analysis}
\subsubsection{Brownian motion on networks}
The simulations are tested on synthetic networks, as well as network structures extracted from images of the ER. For comparing particle distributions on the triskelion network (Fig.~\ref{fig:simMSD}b), $10^7$ particles were simulated with timesteps of $\Delta t D /\ell^2 = 0.01$, for $80$ steps. All particles were started on the central node.

The synthetic honeycomb network in Fig.~\ref{fig:simMSD}c was created by cropping a standard honeycomb lattice with $N=15$ cells in each dimension to a circle of radius $1$. All reported length units for simulations on synthetic networks are normalized by the single edge length $\ell$ in the network. The decimated honeycomb structure was obtained by removing $30\%$ of all network edges, randomly selected in such a way that the network maintains a single connected component. 
While all the example networks used in this study are two-dimensional, the simulations and unraveling code also applies to 3D networks.

Brownian simulations for the MSD comparison were run with $100$ particles, using dimensionless timesteps of $\Delta t D/\ell^2 = 0.01$, for up to $N_t = 3 \times 10^4$ time steps. The mean squared dispacement (MSD) is computed as $\text{MSD} = \left<|\vec{x}(t) - \vec{x}(0)|^2\right>$, where the average is done over non-overlapping time-windows of a single trajectory and over all particle trajectories. 

\subsubsection{Trajectory unraveling analysis}
The algorithm for unraveling observed trajectories on a network to sample the corresponding trajectories on an infinite line is described in Sec.~\ref{sec:unravel}. We perform the unraveling procedure with $20$ values of diffusivity in the range $0.2-4$ (compared to the simulated value of $D=1$). Residuals are calculated by interpolating each unravelled MSD curve onto $100$ logarithmically spaced time-points and then applying Eq.~\ref{eq:residuals}. The self-consistent diffusivity is then found by estimating the value of $D$ that yields a local minimum in the normalized residuals curve (eg: Fig.~\ref{fig:unravelsim}c) via a cubic spline interpolation.

\subsection{Simulation of fractional Langevin motion on networks}
\label{app:fLM}
The long-tailed memory kernel of fractional Langevin motion implies that individual steps of such particles are not Markovian, necessitating an alternate approach to the simulation. We take the approach of assuming that individual particles are undergoing fLM in two dimensions, but that rapid reflections from the walls of a tubule prevent their perpendicular motion. Our overall approach is analogous to that described previously~\cite{jeon2010fractional} for simulating the fractional Langevin equation under confinement, in the limit of no inertia (as appropriate for highly viscous intracellular fluids). 

Specifically, we consider the power-law memory kernel~\cite{weber2010subdiffusive}
\begin{equation}
\begin{split}
K(t-t') = \frac{(2-\alpha)(1-\alpha)}{|t-t'|^{\alpha}}
\end{split}
\end{equation}
and generate 2D stochastic forces $F_x^{(B)}, F_y^{(B)}$ as fractional Gaussian noise~\cite{kroese2013spatial} with time correlations in accordance with the fluctuation dissipation theorem:
\begin{equation*}
\begin{split}	
\left<F^{(B)}_i(t) F^{(B)}_j(t')\right> = \gamma k_bT K(t-t') \delta_{ij},
\end{split}
\end{equation*}
where $\gamma$ is the friction coefficient and $k_bT$ the thermal energy of the particles.
The particle motion obeys the overdamped Langevin equation:
\begin{equation}
\begin{split}
	0 = - \gamma \int_0^t K(t-t') \vec{v}(t') dt' + \vec{F}^{(b)}(t)
\end{split}
\label{eq:fLE}
\end{equation}
where $\vec{v}(t')$ is the instantaneous particle velocity. In the absence of confinement, the MSD for such particles is described by the following power law~\cite{weber2010subdiffusive}:
\begin{equation}
\begin{split}
	\left<x^2\right> & = 2 D_\alpha t^\alpha ,\\
	D_\alpha & =  \frac{k_bT/\gamma}{(2-\alpha)(1-\alpha)\Gamma(\alpha+1)\Gamma(1-\alpha)}
\end{split}
\end{equation}

To propagate forward the particles via discrete time-steps, we take the approach described in Ref.~\cite{deng2009ergodic}. Namely, we discretize the integral in Eq.~\ref{eq:fLE} and solve for the next trial spatial step $\Delta\widehat{\vec{x}}_{n+1}$ based on all prior steps $\Delta \vec{x}_{n+1}$ and the fractional Brownian force $\vec{F}^{(B)}_{n+1}$. 

The trial step $\Delta\widehat{\vec{x}}$ is then projected along the direction of the current edge containing the particle (removing the perpendicular component). If the remaining step takes the particle past a degree-1 node, it is reflected off the node, back along the same edge. If it steps past a degree-3 node, then it is placed randomly on one of the adjacent edges (including the one it came from). For our simulated networks, edges are evenly distributed around each such junction, so that multiple reflections in the tight junctional space are expected to be equally likely to bounce the particle to any edge. Application of this algorithm to more complex structures, including bent degree-2 nodes and degree-3 nodes with uneven edge distribution would require further refinement of this algorithm (left to future work) to account for biased probabilities of entering subsequent edges.

The final saved step $\Delta{\vec{x}}$ is then computed as the two-dimensional difference between the final particle position and its prior position at the start of the step. This saved step is used in the computation of all future particle displacements, allowing the particle to maintain a memory of its prior speed and direction of motion.

We note that this simulation approach gives the expected mean squared displacement of $2D_\alpha t^\alpha$ when particles are placed along a single long edge or along a fully connected honeycomb lattice.

\subsection{Experimental methods}
\subsubsection{Cell culture, plating, and transfection}
COS7 cells were purchased from ATCC and maintained in phenol red-free Dulbecco’s modified Eagle medium supplemented with 10\% (v/v) FBS, 2mM L-glutamine, 100U/ml penicillin and $100\mu\text{g/ml}$ streptomycin at 37\degree C and 5\% CO\textsubscript{2}. All experiments were performed within 40 passages of the initial thaw, and passaging was performed using phenol red-free trypsin (Corning).

High tolerance, 25mm Number 1.5 coverslips were purchased from Warner scientific and precleaned with a modified version of a previously described protocol \cite{sengupta2013quantifying}. Briefly, the coverslips were sonicated for 12 hours in 0.1\% Hellmenex™ (Sigma), followed by five washes in 300ml of distilled water, followed by an additional 12 hour sonication in distilled water and an additional round of five washes. Coverslips were then ethanol sterilized in 200 proof ethanol and allowed to air dry in a clean tissue culture hood. After cleaning, coverslips were stored in an airtight container until ready for use.

Coverslips were pre-coated with $500\mu\text{g/ml}$ phenol red-free matrigel (Corning). Cells were seeded to achieve ~60\% confluency at the time of imaging. Transfections were performed after letting the cells adhere to the coated glass for at least 12 hours, using Fugene6 (Promega) according to the manufacturer's protocol. Each coverslip was transfected with 750 ng of PrSS-mEmerald-KDEL to label the ER structure in the 488 channel, and with 250ng of HaloTag-Sec61b-TA for tracking on the second camera. The HaloTag construct is a minimal targeting domain from Sec61b fused with a flexible linker and the HaloTag, as this construct has been shown to be biochemically inert within the ER\cite{nixon2016increased}.

\subsubsection{Labeling conditions}
Immediately before imaging, each sample was labeled with 10nM PA-JF646 \cite{grimm2016bright}  in OptiMEM (ThermoFisher) for 1 minute. Following staining, cells were immediately washed at least 5 times with 10 ml of PBS under continuous aspiration, taking care not to let the cells contact the air. This was followed by an additional wash in 10 ml of pre-warmed, complete medium. Samples were then left 10 minutes in an incubator to let the cells recover before moving immediately to the microscope. 

\subsubsection{Microscope and imaging conditions}
Dual-color imaging was performed using a customized inverted Nikon Ti-E microscope outfitted with a live imaging stage to maintain temperature, CO\textsubscript{2} level, and relative humidity during imaging (Tokai Hit). The sample was illuminated with two fiber-coupled 488nm and 642nm lasers (Agilent Technologies) introduced into the system with a conventional rear-mount TIRF illuminator. Imaging was performed with the angle of incidence of excitation light manually adjusted beneath the critical angle as needed to produce the most even illumination possible in the ER. If necessary, a small amount of 405nm light was introduced to increase the rate of photoconversion in sparsely labeled cells, but in practice this was rarely needed. The 488 illumination power was kept beneath 100 $\mu$W total in the back aperture, in order to avoid undesirable activation of the photoconvertible dye.

Fluorescence emission light was collected with 100x $\alpha$-Plan-Apochromat 1.49 NA oil objective (Nikon Instruments). Emission light was split into dual paths using a 565LP or 585LP dichroic mirror in a MultiCam optical splitter (Cairn Research), and the two channels were cleaned up by passing the light through 525/50 and 647LP filters (Chroma) placed before the camera. Final signal was focused back onto synchronized dual iXon3 electron mutliplying charged coupled device cameras (EM-CCD, DU-897; Andor Technology). Imaging was performed with 5ms exposure times for 60-90 seconds at a time, and the timing of each frame was monitored using an oscilloscope directly coupled into the system (mean frame rate $\approx95$Hz). 

\subsection{Image analysis for experimental data}
\subsubsection{Image preparation and preprocessing}
The ER is relatively stable at the level of diffraction-limited imaging over the time scale of a second~\cite{friedman2010er}. Thus, we performed 10 frame (110 msec) median filter on the channel for the ER structure to increase the signal to noise and minimize the necessary 488 radiation. Filtering was performed for every frame, but for segmentation and downstream analysis, the structure was analyzed approximately every second (100 frames). 

\subsubsection{Segmentation}
Segmentation and analysis of the tubular ER structure was performed using a two-step process. First, the intensity in the filtered image was made uniform through time with a simple ratio bleach correction. The location of the ER within the image was identified using an interactive pixel classification workflow in Ilastik~\cite{berg2019ilastik}. 
Once the predictions for the ER label were judged to be of sufficiently good quality, the image was made binary using a simple threshold on the label probability exported from ilastik. 

\subsubsection{Extracting network structure}
For each binary segmented image of the ER structure, we skeletonize using the bwmorph subroutine in MATLAB, which also identifies junction pixels. Neighboring junction pixels are grouped into a single node located at their center of mass. We then use the bwtraceboundary subroutine to trace out the skeletonized edges starting from each node, until a neighboring node is reached. This is repeated until the full skeleton has been categorized into nodes and edges connecting specific node pairs. The path of the edge between each pair of connected node is smoothed using cubic splines. Network construction and manipulations is carried out using Matlab code provided at: \url{https://github.com/lenafabr/networktools}.

\subsubsection{Particle trajectories}
Single molecule localization and tracking was performed using the TrackMate plugin in Fiji \cite{tinevez2017trackmate,jaqaman2008robust}. Linking parameters were experimentally selected for each data set to minimize visible linkage artifacts as determined by eye. Datasets were then projected onto the simultaneously collected structure of the ER and manually curated to remove trajectory linkages that were close in 2D but far from one another in the underlying organelle.

For each network structure, obtained at $\approx 1.08$ sec ($100$ frame) intervals, we identify particle trajectory segments that fall within $\pm 100$ frames of the timepoint corresponding to that structure. Those trajectories are then projected onto the network structure by finding the nearest point along the edge paths to each particle position. Because the time intervals surrounding each network structure overlap with each other, each trajectory segment enters the analysis twice by projection onto the structure before and after its time-point.

Whenever the original particle position is more than $2$px ($\sim 0.3\mu$m) away from the nearest network edge, the projected point is removed and the trajectory is broken into separate segments. 
 The trajectories are also broken whenever a  particle appears to step onto a non-adjacent edge (bypassing more than one node) within a single time-step.

The projection step allows us to re-express a particle trajectory in terms of the edge index and position along the edge contour at each time step. Projected trajectories of length at least $10$ time-steps were kept for analysis and 
 unravelled using the same algorithm as for simulated data. An average of 380 projected trajectories per cell, with average length 48 timesteps ($0.51$ sec) were analyzed.
 
 \section{Effect of motion around tubule circumference}
 \label{app:circum}
 
 \begin{figure}
 	\includegraphics[width=0.45\textwidth]{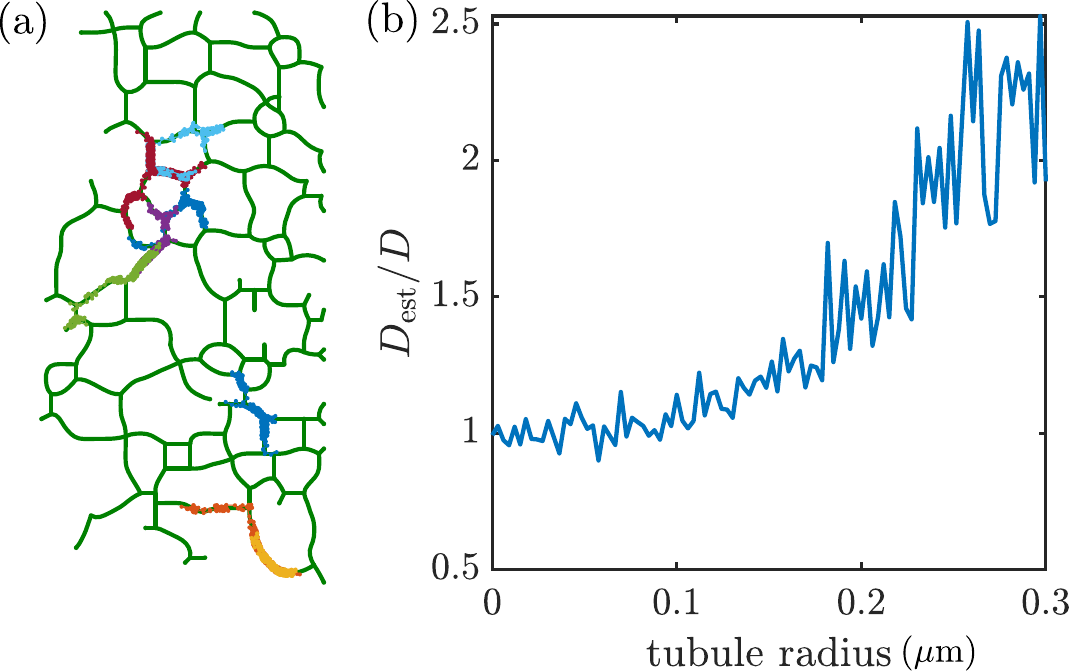}
 	\caption{Effect of particle motion around tubule circumference on diffusivity estimates. (a) Sample trajectories of particles on tubules of radius $100$nm. (b) Estimated diffusivity $D_\text{est}$ normalized by simulated diffusivity $D$, for particles moving along the  network in (a), with different tubule radii. Particles are placed at an arbitrary position around the circumference of the tubule at each step.
 	}
 	\label{fig:simMSDcircum}
 \end{figure}

 Our simulations and data analysis rely on the assumption that the edges of the network are very narrow, and can be treated as effectively one-dimensional. However, in a realistic ER tubular network, membrane proteins are confined to the circumference of the tubule and may thus move perpendicular to the edge axis, as well as along the edge. Typical ER tubules have a radius of approximately $50$nm~\cite{schroeder2019dynamic}. 
 
 Here, we use simulations to examine how displacements along the tubules' circumference (orthogonal to the edge paths) might impact the diffusivity estimation. In particular, we assume tubules are narrow enough that particles can explore the entire circumference over an individual time-step. This is a valid assumption for ER membrane proteins, with approximate diffusivity $1.5\mu\text{m}^2/\text{s}$, where the timescale to cover the circumference is approximately $0.008$sec. Thus, we assume at each step that the circumferential position of the particle (around the cross-section perpendicular to the edge path) is selected uniformly, with no correlation to the previous steps.
 
  When the particle position is projected in two dimensions, the displacement orthogonal to the edge axis is in the range $-r \leq x \leq r$, where $r$ is the tubule radius. This displacement obeys the cumulative distribution function:
\begin{equation}\label{noise_distribution}
    f(x) = \frac{1}{2}+\frac{1}{\pi}\arcsin{\frac{x}{r}}% \begin{cases}
\end{equation}
We simulated particle trajectories on the network shown in Fig.~\ref{fig:simMSDcircum}a for tubule radii $r$ up to $300$nm. Each of the trajectories was then analyzed in a manner analogous to experimental data, by projecting the particle onto the nearest point along the network edges. After applying the unraveling procedure, we see that the estimated diffusivity is not substantially affected by the circumferential displacements for tubules of radius below $100$nm. Tubules in the peripheral ER fall well within this criterion, and we thus expect our assumption of infinitely narrow tubules to not yield substantial bias in the analysis.

\end{document}